%% file: main.tex
\documentclass[letterpaper,twocolumn,10pt]{article}
\usepackage{times}

\usepackage{usenix2019_v3}
\usepackage{tikz}
\usepackage{threeparttable}
\usepackage{soul}
\usepackage{colortbl}
\usepackage{multirow}
\usepackage{tabularx}
\newcolumntype{Y}{>{\centering\arraybackslash}X}
\usepackage{amsmath,pifont, mathtools}
\usepackage{amsmath}
\usepackage[overload]{empheq}
\usepackage{nccmath}
\usepackage{algorithm}
\usepackage{cases}
\usepackage{bbding}
\usepackage{graphicx}
\usepackage{pifont}
\usepackage{wasysym}
\usepackage{algorithmicx}
\usepackage[noend]{algpseudocode}
\usepackage{mathtools}
\usepackage{float}
\usepackage{authblk}
\usepackage{subcaption}
\usepackage{adjustbox}
\usepackage{makecell}%
\usepackage{tikz}
\makeatletter
\newcommand*{\rom}[1]{\expandafter\@slowromancap\romannumeral #1@}
\makeatother
\newcommand{\tbg}{\cellcolor[HTML]{D0D0D0}}
\newcommand*\circled[1]{\tikz[baseline=(char.base)]{
            \node[shape=circle,draw,inner sep=0.5pt] (char) {#1};}}

\usepackage[english]{babel} % for use of English
\newcommand{\cmark}{\ding{51}}%
\newcommand{\xmark}{\ding{55}}%

\usepackage{booktabs}
\usepackage{hyperref}

\usepackage{cleveref}
\crefname{section}{}{\S\S}
\crefdefaultlabelformat{\S#2#1#3}

\usepackage{xspace}
\usepackage{comment}
\usepackage[normalem]{ulem}
\usepackage{enumitem}
\setlist{nolistsep}

\usepackage{listings}
\usepackage{color}
\usepackage{caption}
\usepackage{textcomp}

\definecolor{dkgreen}{rgb}{0,0.6,0}
\definecolor{gray}{rgb}{0.5,0.5,0.5}
\definecolor{mauve}{rgb}{0.58,0,0.82}
\DeclareCaptionLabelFormat{andtable}{#1~#2  \&  \tablename~\thetable}
\usepackage{balance}

\newcommand{\ljm}[1]{{\color{blue} #1}}

\newcommand{\sys}{Aryl\xspace}
\newcommand{\sysP}{Aryl+TunedJobs\xspace}

\graphicspath{{figures/}}
\hypersetup{draft}
\hypersetup{final}
\begin{document}

%make title bold and 14 pt font (Latex default is non-bold, 16 pt)
\title{\sys: An Elastic Cluster Scheduler for Deep Learning}
% \title{\sys: Elastic Cluster Scheduling for Deep Learning}
\date{}
% \author{
% {\rm Jiamin Li$^{1, 2}$,} \and 
% {\rm Hong Xu$^{3}$,} \and
% {\rm  Yibo Zhu$^{1}$,} \and 
% {\rm Zherui Liu$^{1}$,} \and 
% {\rm Chuanxiong Guo$^{1}$,} \and 
% {\rm Cong Wang$^{2}$,} \\
% $^{1}$ByteDance, $^{2}$City University of Hong Kong, $^{3}$The Chinese University of Hong Kong
% }
\author[1, 2]{Jiamin Li}
\author[3]{Hong Xu}
\author[1]{Yibo Zhu}
\author[1]{Zherui Liu}
\author[1]{Chuanxiong Guo}
\author[2]{Cong Wang}
\affil[1]{ByteDance}
\affil[2]{City University of Hong Kong}
\affil[3]{The Chinese University of Hong Kong}
\maketitle  
\thispagestyle{empty}
%\subtitle{Extended Abstract}

%for single author (just remove % characters)
% \author{
% {\rm Paper \#97, 12 pages with 2-page references}
% Your Institution
% \and
% {\rm Second Name}\\
% Second Institution
% copy the following lines to add more authors
% \and
% {\rm Name}\\
% Name Institution
% } % end author

% \author{Paper \#88, 12 pages}
% \affiliation{%
  % \institution{}
% }

% \author{Technical report}

\input{abstract}

% \vspace{-15mm}
\pagestyle{plain}

\input{introduction}
\input{motivation}
\input{design}
\input{loaning}
\input{elastic}
\input{implementation}
\input{evaluation}
\input{discussion}
\input{related}

% \vspace{-2mm}
\section{Conclusion}
% \vspace{-2mm}

We have presented \sys, an elastic GPU cluster scheduler for deep learning.
The key idea is to exploit cluster-level elasticity by loaning idle inferences servers for training, and job-level elasticity by scaling jobs to better utilize the dynamic resource pool.
In designing and evaluating \sys, we have addressed new challenges in cluster management,
by introducing heuristics to reduce job preemption cost due to loan-reclaiming,
and to minimize job completion time when elastic jobs are presented.
We plan to extend \sys to support training over heterogeneous GPUs, and to investigate information-agnostic
scheduling without knowing jobs' running time a priori.

% \begin{comment}
\section{Acknowledgment}
The project is supported in part by CUHK grants 4055138, 4937007, 4937008, 5501329, 5501517.
% \end{comment}
\clearpage
\bibliographystyle{plain}
\bibliography{main}
\clearpage
\input{appendix}
% \section*{Link to Appendices}
% The appendices are provided as a separate document here: \url{https://www.dropbox.com/s/kltb7gup6p823n8/appendix_main.pdf?dl=0}
\end{document}

%% file: abstract.tex
\begin{abstract}

Companies build separate training and inference GPU clusters for deep learning, and use separate schedulers to manage them.
This leads to problems for both training and inference: inference clusters have low GPU utilization when the traffic load is low; training jobs often experience long queueing time due to lack of resources.
We introduce \sys, a new cluster scheduler to address these problems.
\sys introduces \textit{capacity loaning} to loan idle inference GPU servers for training jobs. It further exploits \textit{elastic scaling} that scales a training job's GPU allocation to better utilize loaned resources.
Capacity loaning and elastic scaling create new challenges to cluster management. When the loaned servers need to be returned, we need to minimize the number of job preemptions; when more GPUs become available, we need to allocate them to elastic jobs and minimize the job completion time (JCT).
\sys addresses these combinatorial problems using principled heuristics. It introduces the notion of server preemption cost which it greedily reduces during server reclaiming. It further relies on the JCT reduction value defined for each additional worker for an elastic job to solve the scheduling problem as a multiple-choice knapsack problem.
Prototype implementation on a 64-GPU testbed and large-scale simulation with 15-day traces of over 50,000 production jobs show that \sys brings 1.53x and 1.50x reductions in average queuing time and JCT, and improves cluster usage by up to 26.9\% over the 
cluster scheduler without capacity loaning or elastic scaling.

\end{abstract}

%% file: introduction.tex
%!TEX root = main.tex

\section{Introduction}
\label{sec:introduction}

Recently, Deep Neural Networks (DNNs) have seen wild successes in many
applications \cite{lecun2015deep}. Hyperscale online service providers have
adopted DNN, and build large-scale GPU clusters to accelerate DNN workloads
for both training and inference. GPU cluster scheduling is a fundamental and
critical task to utilize the expensive GPU clusters efficiently, by optimizing
the job resource allocation and task placement.

It is common practice today to separately build and manage two types of GPU clusters,
one for training and one for inference.
This is because, for the same model, inference requires less computation and
GPU memory than training and is less likely to utilize the numerous cores of training GPU
\cite{park2018deep, crankshaw2017clipper, narayanan2018accelerating}.
Inference clusters usually use weaker GPUs, like Nvidia T4, with a fraction of the resources of the training GPUs, e.g., Nvidia V100 and A100.
% The schedulers for training and inference are also designed with different
% objectives and constraints \cite{xiao2018gandiva,258957,10.1145/3341301.3359658}.

% For training clusters, their schedulers usually
% aim to minimize the job completion time (JCT) \cite{gandiva, tiresias,
% antman}, while inference
% schedulers optimize throughput and GPU utilization given the latency target
% \cite{nexus}.

This separation creates problems for both sides
(\cref{sec:motivation}).
Our observations are based on experiences of operating production clusters
with $O$(10K) GPUs for training and even more for inference.
Specifically, inference cluster utilization is usually low ($<$40\%) for an extended period of time due to the diurnal traffic pattern.
At the same time, training jobs experience long queuing before they can
start, with an average of over 3,000s and 95\%ile of almost
10,000s as seen from a 15-day trace with over 50,000 jobs.
The long queuing time is due to both the high cluster utilization and
the GPU resource fragmentation.

To address the above problems, we propose \textit{capacity loaning} to allow the
inference cluster to loan the idle GPU servers during low-traffic periods to run
training jobs, and reclaim them back when inference workloads increase again
(\cref{sec:loaning}). Capacity loaning mitigates both the utilization problem
for inference and queuing problem for training. It is feasible for training jobs
which do not have strict requirements on GPU type. For on-loan servers, we
need to ensure that they are rapidly utilized by training jobs when they become
available.
% We explore two dimensions to facilitate this.
We draw inspiration from \textit{elastic scaling}
\cite{ElasticHorovod:Website,PyTorchElasticLaunch:Website,or2020resource}
for training jobs to better use the on-loan servers (\cref{sec:elastic}).
Elastic scaling enables a running job to scale out or scale in to better utilize the dynamically changing resource pool.
It also helps reduce queueing delay since an elastic job can start with a
small number of workers first and increase its workers when more resources become available.

% to the training cluster, we find that those GPUs
% are often under-utilized, because the
% training cluster may not immediately have enough queuing jobs to be scheduled on to those GPUs.
% Fortunately, an emerging techinique in training, elastic training~\cite{PyTorchElasticLaunch:Website,ElasticHorovod:Website},
% fits this scenario well. Elastic training allows
% an already-running multi-GPU jobs to scale out or scale in following the commands of cluster scheduler.
% Once the training clusters loan GPUs from the inference cluster, elastic training can help jobs flexibly
% leverage the added GPUs.

Capacity loading and elastic scaling create new degrees of freedom for cluster
scheduling.
As we navigate the new design space, we meet several
new challenges that must be addressed before we can reap the benefits.

First, though loaning decisions can be solely made by the inference cluster scheduler
to ensure inference workloads are not affected, reclaiming is more intricate.
When the inference cluster needs to reclaim some on-loan servers, the
training scheduler has to preempt all running jobs on those servers.
Given the high overhead and prolonged running time associated with preemption,
the scheduler must carefully select the servers in order to minimize the total
preemptions.

Second, the job scheduling problem is inherently more complicated with elastic scaling.
Resource allocation has to consider a mix of inelastic jobs with fixed
demand and elastic jobs
with \textit{variable} demand.
We show that classical scheduling policies such as
shortest job first (SJF) no longer works well with elasticity, and finding the
JCT-optimal solution for merely two jobs is difficult.
Given the allocation results, the scheduler still needs to determine the
worker-server placement to minimize fragmentation, where servers are now
heterogeneous with different GPUs because of capacity loaning.

Our key intuition in solving these challenges is to prioritize the minimum
resources needed by each job over the elastic demand, and to prioritize the
dedicated training servers over the on-loan inference servers.
This makes sense because the minimum demand of an elastic job is equivalent to
an inelastic job to which not allocating resources is detrimental, but the
elastic part can be fulfilled later without stalling the job.

Our solution therefore exhibits a two-phase structure following the above
intuition.
For reclaiming, we first kill the elastic workers running on on-loan servers since stopping them does not lead to any job-level preemption.
%  (\cref{sec:placement_design}).
% This is made easy by our placement that separates elastic workers on a subset of on-loan servers already.
When preemption becomes inevitable, we characterize the problem as a knapsack
problem with dependent item values \cite{mougouei2017integer} and develop an efficient heuristic to solve it (\cref{sec:design_reclaim}).

For resource allocation, we first allocate for both inelastic
jobs and elastic jobs' base demand, with the aim of launching as many
jobs as possible.
We then scale out the scheduled elastic jobs if resources permit.
% \ljm{Update}
The first phase can be solved using SJF to reduce queuing time and
the second phase is formulated as a multiple-choice knapsack problem \cite{sinha1979multiple}
to minimize running time, which in practice can often be solved using
dynamic programming (\cref{sec:greedy_allocation}).
We then tackle the placement problem by placing the inelastic jobs to training servers, and elastic jobs to on-loan servers as much as feasible.
Jobs are ordered based on the best-fit-decreasing policy to address the
bin packing nature \cite{CoffmanJr.2013} and minimize fragmentation
(\cref{sec:placement_design}).

Putting everything together,
we design (\cref{sec:design}--\cref{sec:design_elastic}), implement (\cref{sec:implementation}), and evaluate (\cref{sec:evaluation}) \sys, a new cluster scheduler that realizes capacity loaning with elastic scaling.
\sys has an orchestrator that manages capacity loaning by executing instructions
from the inference scheduler on when and how much to loan or reclaim, and
by deciding which on-loan servers to return for reclaiming.
Then a job scheduler periodically determines allocation and placement, and
scales new and existing elastic jobs in response to the resource and job dynamics.
To be pragmatic, \sys considers elastic scaling only for large DNNs whose training throughput scales well in our experiments.

The results of \sys are promising (\cref{sec:evaluation}).
We build a high-fidelity simulator, and replay a 15-day job trace collected from 3,544 training GPUs and 4,160 inference GPUs.
We find that compared to a FIFO scheduler, \sys can reduce the average and 95\%ile JCT by up to 1.50x and
1.47x, respectively, and improve GPU usage by 26.9\%. 
% Compared with existing elastic job schedulers, 
In terms of job scheduling, \sys also outperforms state-of-the-art Pollux by 1.32x and 1.37x in median JCT and 95\%ile JCT when elastic jobs occupy 36\% training resources. 

% In a small cluster with 64 GPUs, we replay a scaled-down job trace.
% and find that the average JCT of training jobs is reduced by 1.22x and the 95th percentile JCT by 1.13x.

We summarize our contributions as follows.
\begin{itemize}[leftmargin=4mm]
	\item We report problems of separate management of training
	and inference clusters, i.e. low utilization in the inference cluster and
	long queuing time in the training cluster, measured from production GPU clusters.

	\item We propose cluster-level capacity loaning and job-level elastic scaling,
	two new control knobs for cluster scheduling to address the above problems.

	\item We study the resulting cluster scheduling
	problems, develop a key intuition to prioritize the minimum resources
	needed by each job to address elasticity, and use a principled approach to
	characterize and solve each problem.

  \item We design and implement \sys, a novel cluster scheduler that
  integrates our solutions. \sys works with existing resource
  management frameworks and is ready for deployment. Evaluation using
  testbed experiments and
  large-scale simulations validates its superior performance.

    % To our knowledge, it is the first achieving those    and showing
    % significant benefits for for real-life DNN workloads.

    % \item We analyze the problems of inference cluster reclaiming resources from training. We design a heuristic strategy that
    % minimizes wasted resources during reclaim.

    % \item We then show the difficulties in scheduling elastic training jobs. We propose a 2-phase algorithm to address this problem.
\end{itemize}

%% file: motivation.tex
%!TEX root = main.tex
\section{Motivation}
\label{sec:motivation}

%We start by presenting the background of deep learning clusters and motivation of \sys.

\subsection{Why Capacity Loaning? }
% \subsection{Why Capacity Loaning? An Empirical Study of DL Cluster Workloads }
% \subsection{DNN Cluster Characterization}
\label{sec:loaning}
Large GPU clusters are built to accommodate inference and training
workloads with distinct requirements.
Inference jobs are latency-sensitive since they are customer-facing
\cite{crankshaw2017clipper, park2018deep}.
Training jobs are much more resource-heavy and run for an extended period
of time.
Thus they emphasize on job completion times instead.
Operators usually deploy separate clusters with different GPU for
training and inference, and manage them independently to minimize interference.
Our production environment, for example, mainly uses Tesla V100 in the
training cluster and T4 in the inference cluster.
Job traces show that this practice leads to low utilization of inference resources and
sub-optimal performance for training jobs.

\noindent\textbf{Inefficient inference resource utilization.}
Similar to many web services~\cite{LWAT11}, the inference cluster is
overprovisioned in order to handle the peak traffic.
Inevitably, its resources are often underutilized due to
the dynamic inference requests generated by customers.

\begin{figure}[t]
    \centering
    \includegraphics[width=0.90\linewidth]{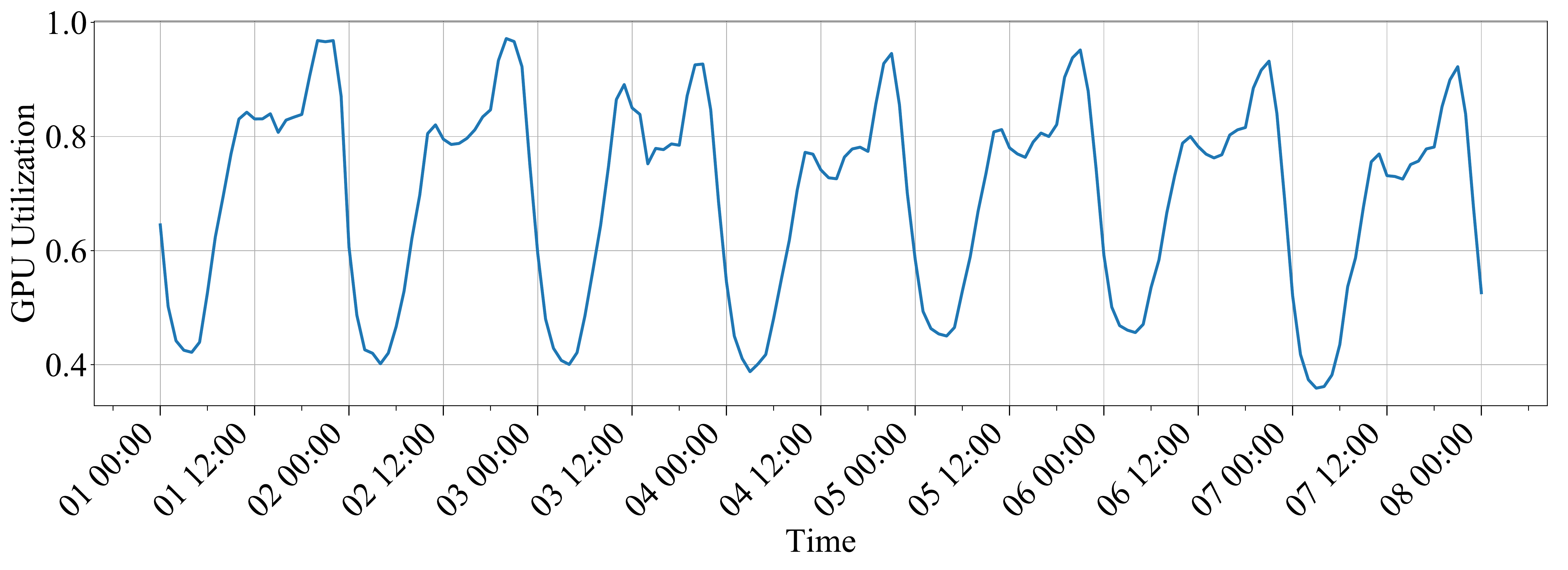}
    \vspace{-4mm}
    \caption{Inference cluster GPU utilization, i.e. fraction of GPUs serving at least one
    request, in our inference cluster.
    The measurement spans one week's time from Oct 1 to Oct 7, 2020.
    The cluster has about 4,000 GPUs.
    In peak hours GPU utilization approaches 95\%.}
    \vspace{-2mm}
    \label{fig:inference_util_diurnal}
\end{figure}

We plot the GPU utilization in one of our inference clusters with 5-minute
intervals for one week's time in Figure~\ref{fig:inference_util_diurnal}.
Utilization is defined as the fraction of GPUs occupied by at
least one inference job.
We observe a clear diurnal pattern: peak traffic lasts about
four hours at night, and demand trough occurs before dawn.
The peak-to-trough ratio is $\sim$2.2 within a day, and the average utilization
is $\sim$65\%, both implying that there are abundant resources to be exploited
in the inference cluster for extended periods of time.

\noindent\textbf{Long queuing time for training jobs.}
\begin{figure}[t]
    \centering
    \includegraphics[width=0.90\linewidth]{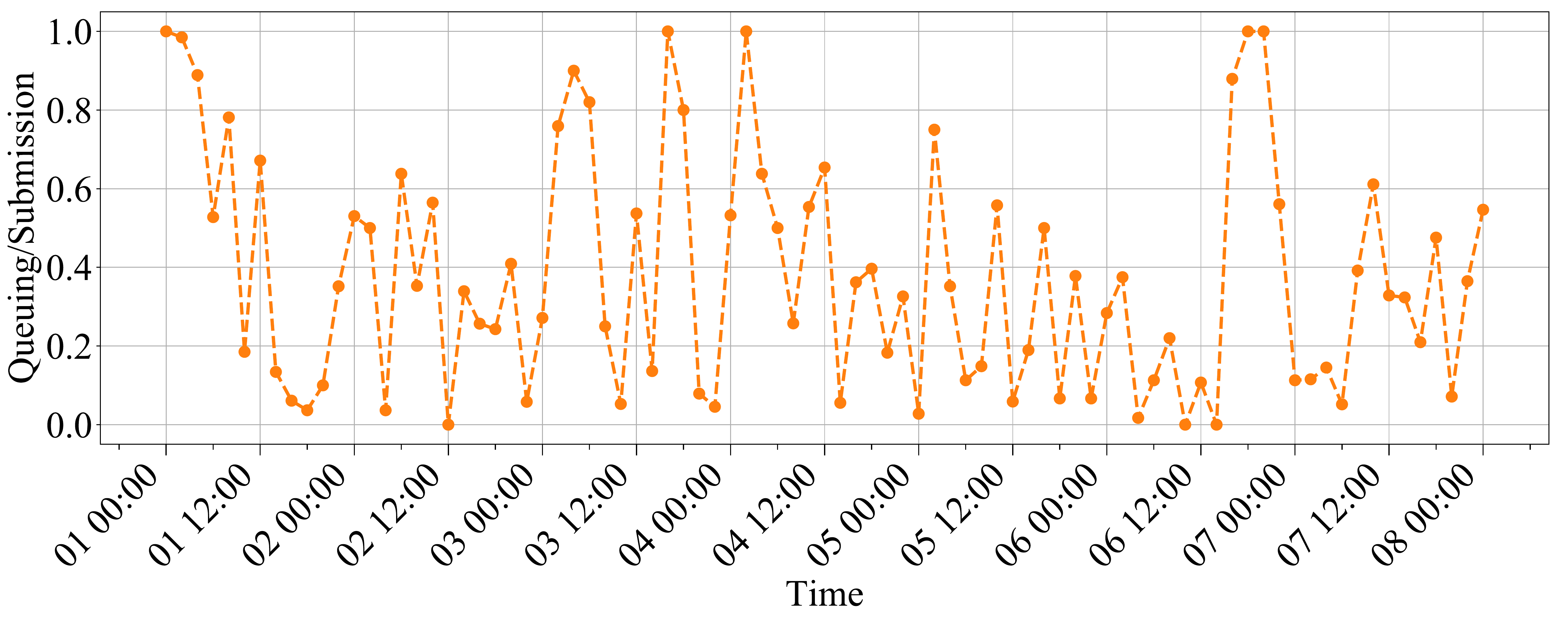}
    \vspace{-4mm}
    \caption{The fraction of queuing jobs among all the newly-submitted jobs in
    each hour in our training cluster for one week's time.
    A job suffers queuing time when the scheduler fails to satisfy its
    resource demand in the first try.
    If the ratio is high, it means most of the jobs submitted in that hour is
    queued.
    The cluster has $\sim$3,500 GPUs and the average utilization is
    82\%.}
    \vspace{-6mm}
    \label{fig:training_queue_util}
\end{figure}
Turning to the training cluster,
a salient observation we make is that many training jobs experience long
queuing time before they can be dispatched with enough resources.
Figure~\ref{fig:training_queue_util} depicts the hourly queuing job ratio
in our training cluster for the same week as in Figure~\ref{fig:inference_util_diurnal}.
%although there are always idle resources in the cluster (average
%utilization is 82\%),
A significant fraction of jobs (as high as 100\%) still have to wait for resources
from time to time. The average queuing time is longer than 3,000 seconds and certainly
non-negligible.
% there are cases when our internal users have to open on-call tickets just to
resolve their queuing delay.

The long queuing time is not only due to lack of resources. In fact, the average
GPU utilization across the same period of time is 82\%, which means there are
often idle GPUs. The dynamic training demand certainly also contributes to the
long queuing time. In addition, training demand does not exhibit a clear pattern for prediction.

% State our main idea here
\noindent\textbf{Capacity loaning.}
We propose to exploit the unused inference resources in demand trough to run
training jobs temporarily, i.e. loaning inference capacity for training.
It mitigates both above problems at the same time:
The inference cluster is better utilized, and training jobs have more
resources to help reduce queuing time.
The on-loan capacity can be reclaimed dynamically in case the inference traffic
spikes to ensure quality of service.

Though training jobs typically request specific GPUs, we find that up to 21\% of jobs in our production traces do not do so and can work with any GPU types.
\sys can launch these jobs on the loaned inference servers rather than waiting for training servers.
To ensure feasibility, we may need to adjust the batch size of the training job so that the models and the intermediate data can fit into the smaller inference GPU memory.
This is straightforward since we know the GPU memory differences; more details can be found in \Cref{app:v100tot4}.

Another more aggressive way to exploit the borrowed servers is to run a training job on heterogeneous GPUs, i.e. using both training and inference GPUs (e.g. V100 and T4).
Heterogeneous training further improves the scheduling flexibility with more performance gains potentially.
However, it requires delicate systems and algorithm support to work well, since the workers have to adopt different hyperparameter settings and inherently make progress at different paces \cite{park2020hetpipe, chen2020semi, NHPS19, yi2020optimizing}.
Given that heterogeneous training remains an active research topic,
% by default our training system disables it for all jobs, and only a small fraction of the jobs in our traces are able to support it with explicit request.
our production training system only provides experimental support for it at the moment.
\sys's design does not depend on it, and we evaluate the effect of heterogeneous training on \sys in \cref{subsec:simulation} when it is enabled for a small fraction of our jobs with non-ideal performance.

\subsection{Elastic Scaling for the Full Potential}
% \subsection{Reaping the Full Potential of Capacity Loaning}
\label{sec:elastic}

% With loaning the capacity of the training cluster is constantly changing.
% To further exploit the inference
To better cope with the constantly changing cluster capacity and further exploit the loaned inference resources, \sys considers \textit{elastic scaling}.
Recently, elastic scaling has been introduced into ML frameworks
\cite{ElasticHorovod:Website, PyTorchElasticLaunch:Website, or2020resource} where a job can take a variable number of workers according to resource availability.
One can even adjust the number of workers on-the-fly when the job is running.
% This becomes instrumental in dealing with the changing cluster capacity.

Elastic scaling can greatly facilitate capacity loaning.
With additional resources, training jobs can dynamically scale out to use more workers with more inference GPUs to accelerate training (provided they are running on inference GPUs already).
When the cluster experiences high loads, some jobs could scale in to free some servers.
In addition, when vacating the inference resources so they can be reclaimed, the scaling-in operation reduces the need of completely preempting the jobs which incurs high overheads with checkpointing, re-launching containers, etc.
\begin{figure}[t]
    \centering
    \begin{subfigure}{0.22\textwidth}
        \includegraphics[width=0.95\linewidth]{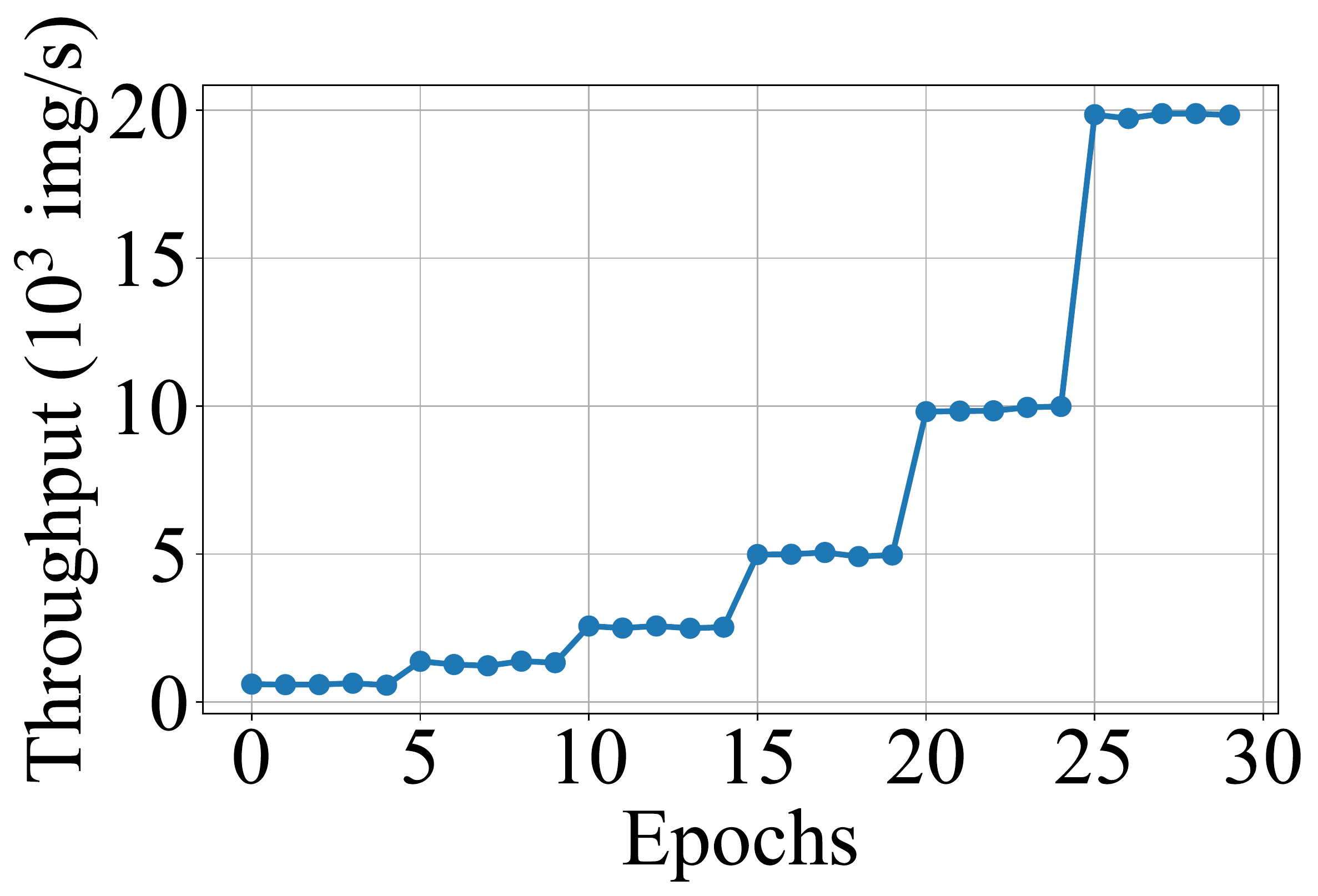}
        \vspace{-2mm}
        \caption{ResNet - Acc: 74.2\%(-0.14\%)}
        \label{fig:resnet_throughput}
    \end{subfigure}%
    \hfill
    \begin{subfigure}{0.22\textwidth}
        \includegraphics[width=0.95\linewidth]{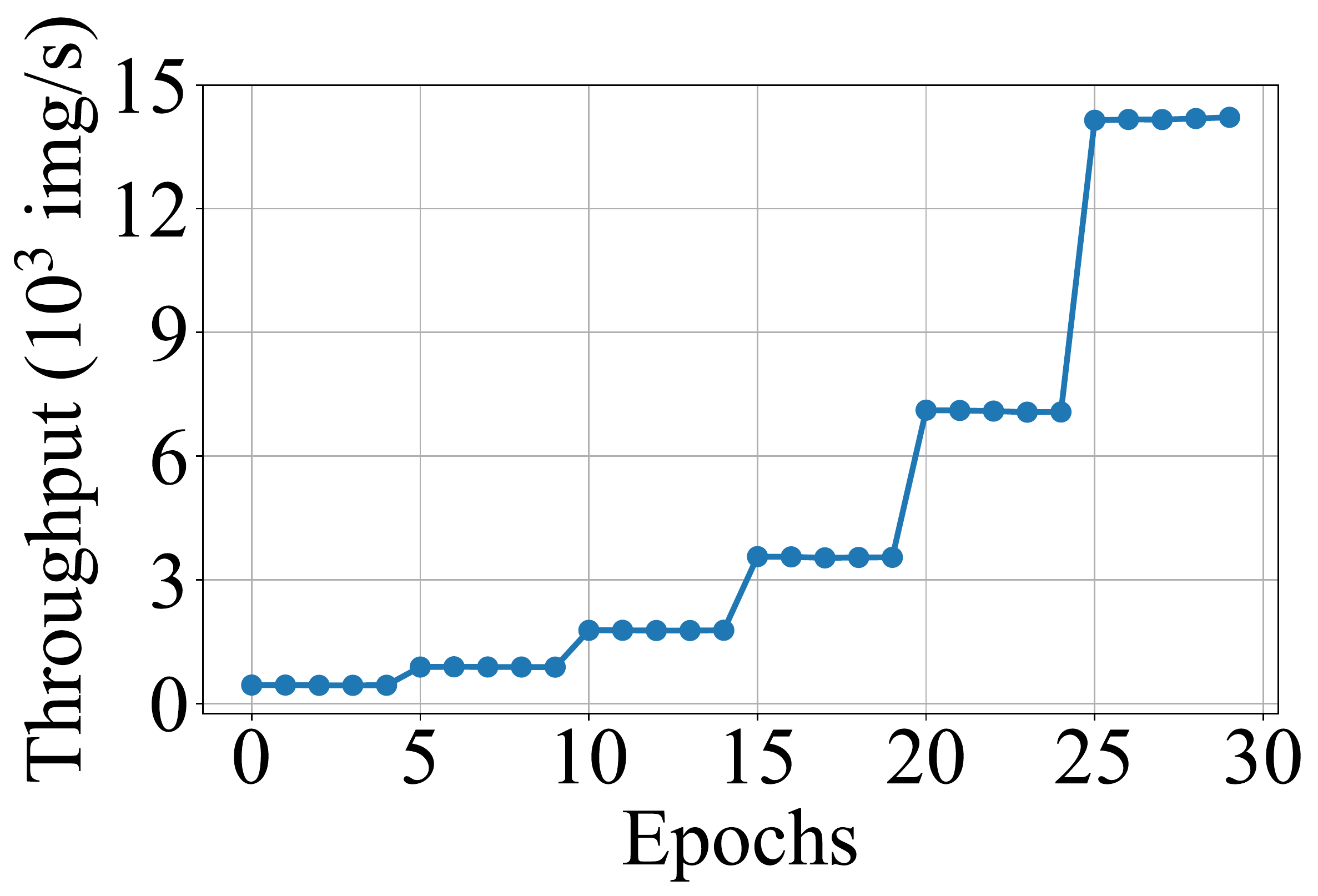}
        \vspace{-2mm}
        \caption{VGG - Acc: 88.5\%(+0.03\%)}
        \label{fig:vgg_throughput}
    \end{subfigure}

    \begin{subfigure}{0.22\textwidth}
        \includegraphics[width=0.95\linewidth]{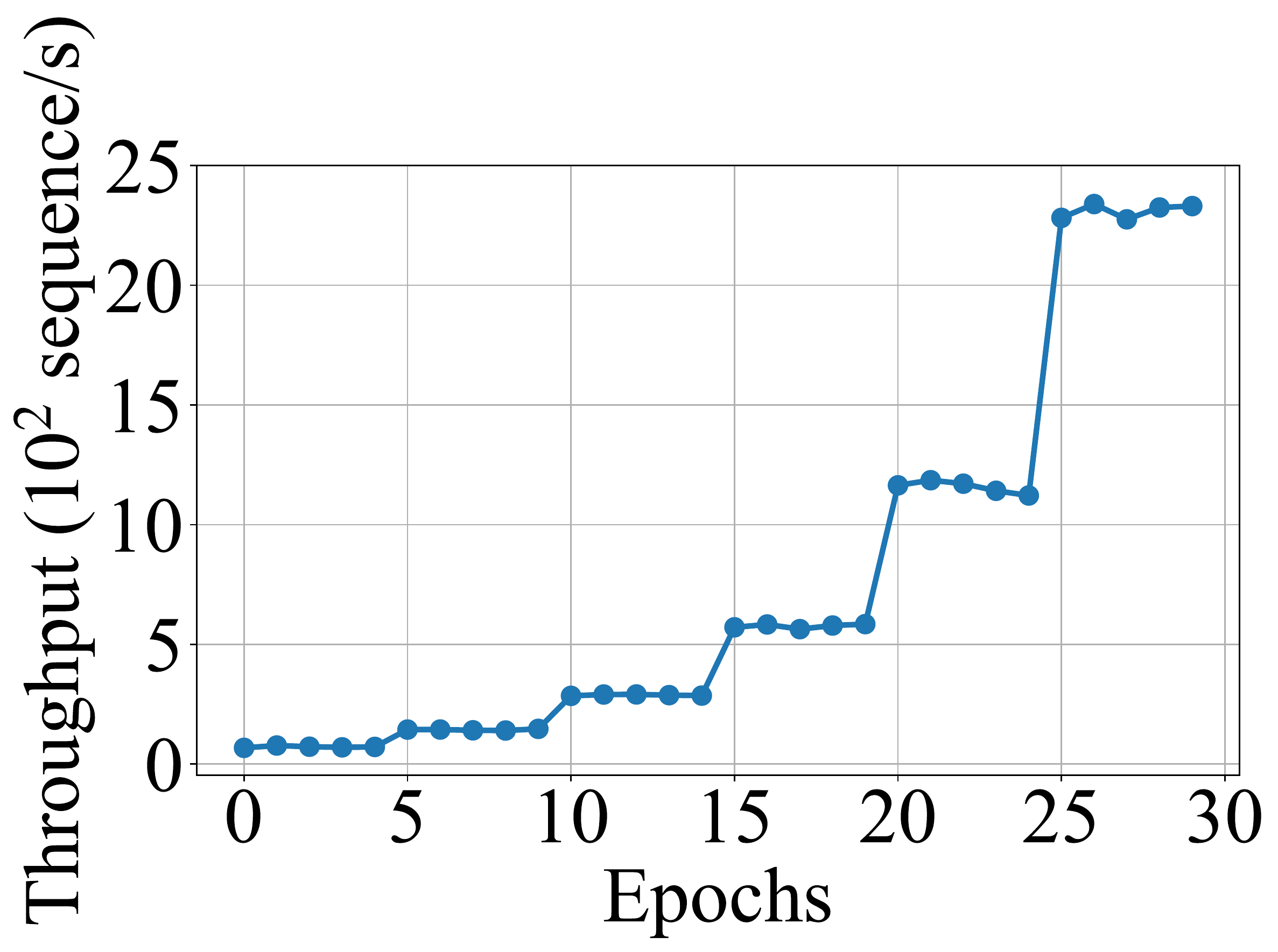}
        \vspace{-2mm}
        \caption{BERT - F1: 88.43(-0.71)}
        \label{fig:bert_throughput}
    \end{subfigure}
    \hfill
    \begin{subfigure}{0.22\textwidth}
        \includegraphics[width=0.95\linewidth]{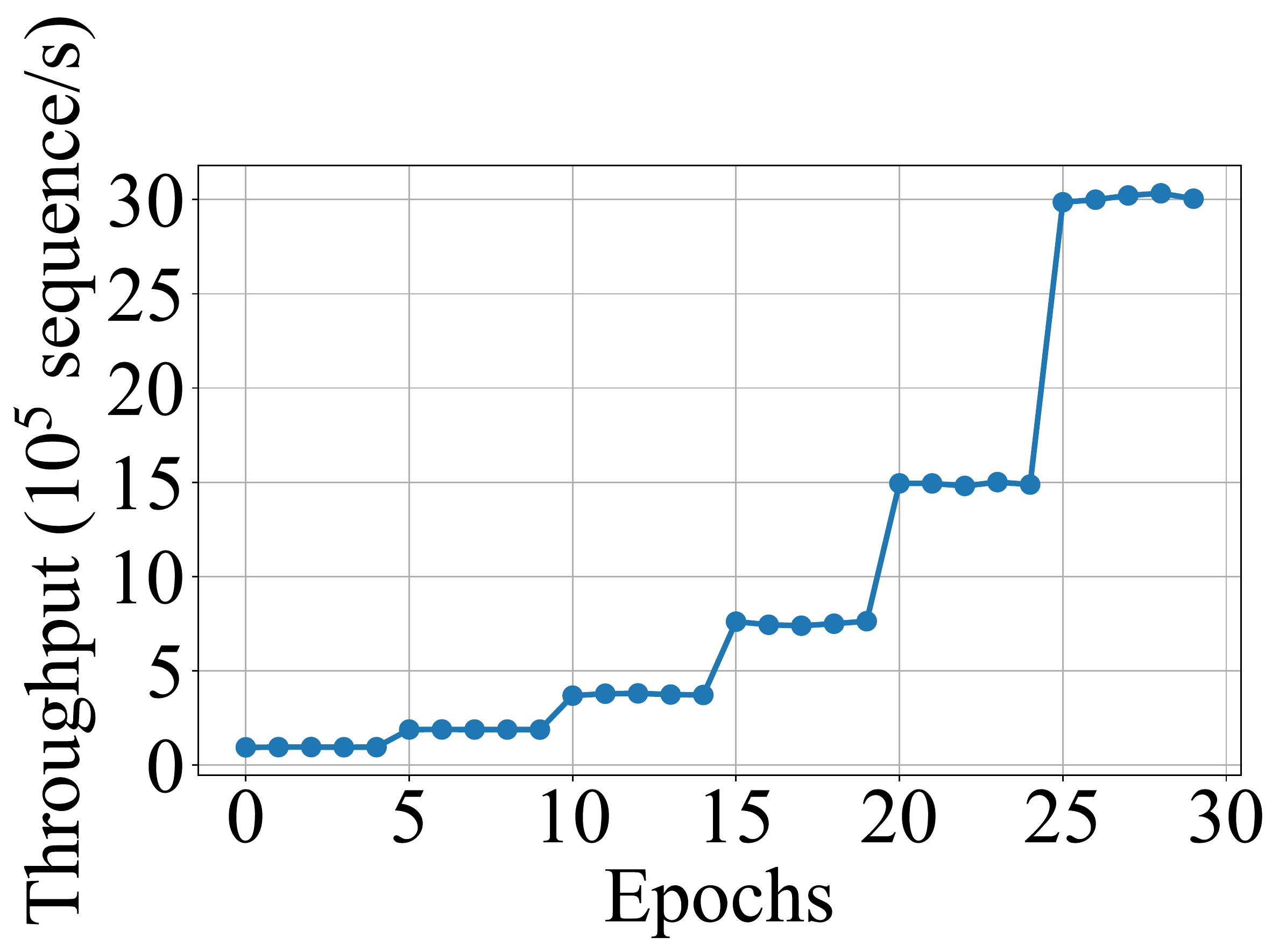}
        \vspace{-2mm}
        \caption{GNMT-16 - BLEU: 25.4(-1.2)}
        \label{fig:gnmt_throughput}
    \end{subfigure}
    \vspace{-2mm}
    \caption{Throughput of four elastic training jobs using Tesla V100 GPUs.
      In our testbed, each server hosts 8 GPUs connected by NVLink.
      Servers use 100G InfiniBand interconnects.
      Each worker container uses 2 GPUs.
    The workers are doubled every 5 epochs, starting from 1 worker. We list the performance
    and the gap with inelastic training in the brackets.}
    \label{fig:elastic_throughput}
    \vspace{-7mm}
\end{figure}
% \ljm{Note: state difficulty, cite to prove the scalability of large dnn jobs, discuss our cluster}
An acute reader might be wondering about the feasibility and benefit of elastic scaling in general.
Indeed, besides the scalability issue of distributed training systems \cite{peng2019generic, JZLY20, yan2015performance, 265013}, when we change the number of workers on-the-fly, the training hyperparameters may have to be updated as well in order to cope with the new setup.
This can be fairly complex: for example, simply keeping the local batch size unchanged and linearly increasing the global batch size may impede the convergence of the model \cite{goyal2017accurate}.

Thus in \sys, elastic scaling is only adopted for jobs that scale well to changing number of workers without updating the local batch size.
Existing studies \cite{jia2018highly, mikami2018massively, chilimbi2014project, mattson2019mlperf,you2019large} show that certain models like ResNet\cite{he2016deep} and BERT\cite{devlin2018bert} satisfy this requirement.
We also find that, as shown in Figure~\ref{fig:elastic_throughput}, ResNet-50\cite{he2016deep}, VGG16\cite{simonyan2014very}, BERT\cite{devlin2018bert}, and GNMT-16\cite{wu2016google} all enjoy good throughput scalability and are well-suited for elastic scaling.
% These four models account for just $\sim$5\% of all jobs in our production traces but require 36\% of training cluster resources with an average running time of 14.2 hours, suggesting ample potential gains using \sys.
Our traces reveal that these large jobs 
account for 36\% of training cluster resources with an average running time of 14.2 hours, suggesting ample potential gains using \sys.
For these jobs \sys also restricts itself to \textit{limited elasticity} where the worker number varies within a range, beyond which more complicated hyperparameter tuning becomes necessary and thus out of scope.

\subsection{Existing Cluster Schedulers}
\label{sec:comparison}
% \ljm{TODO: add a table for comparison}

Much prior work exists on GPU cluster scheduling amid the proliferation of DL workloads.
\sys differs from them mainly in two aspects.

First, capacity loaning represents a new angle to the cluster scheduling
problem few have studied.
Though shared infrastructure is exploited by recent systems 
\cite{verma2015large, lo2015heracles, tumanov2016tetrisched, TWINE, zhu2014prioritymeister, weng2022mlaas},
their focus is to schedule multiple types of workloads in a single
cluster.
\sys instead focuses on virtually loaning resources between two different clusters.
Specifically, it considers the problem of how to reclaim the transient on-loan resources while minimizing its negative impact on training jobs running on them (\cref{sec:design_reclaim}), which has not been considered before.
Further, \sys takes advantage of elasticity of training
jobs to better utilize the dynamic cluster resources. 

%\ljm
{
Second, some recent studies also considered scheduling elastic jobs.
Gandiva \cite{xiao2018gandiva} adopts an opportunistic approach to grow or shrink of number of GPUs used by a job without considering cluster-wide efficiency.
AFS \cite{265013} greedily prioritizes the jobs with the highest 
throughput per GPU.
Pollux \cite{qiao2020pollux} co-optimizes both resource allocation and hyperparameter of DNN jobs to achieve high resource efficiency.

Compared to them, \sys exploits the interplay between elastic scaling and capacity loaning to further improve the performance which has not been explored.
In terms of technical approach, \sys preserves the problem nature of scheduling elastic jobs and treats it as a variant of the knapsack problem, enabling it to make globally good allocation decisions and outperform greedy local heuristics in prior work. 
% These systems simplify the combinatorial nature of the allocation problem by ranking the jobs greedily.
% \sys instead 
% For instance, in placing the jobs, \sys prioritizes elastic jobs to be packed on borrowed servers so that they can just be scaled in instead of being preempted to vacate the servers during reclaiming.
% \sys is also more practical in taking into account realistic constraints in scheduling, such as limited elasticity instead of unbounded elasticity as in most prior work.
Though \sys does not consider tuning hyperparameters, it can be readily integrated into \sys to provide even better performance (\cref{sec:scaling_perf}). 
More details on the differences between \sys and existing schedulers are presented in~\Cref{app:scheduler_detail_comparison} for brevity.
}

%% file: design.tex
\section{Design Overview}
\label{sec:design}
In this section, we describe \sys's overall architecture and the key design questions we need to address.
% Detailed design of cluster-level capacity loaning and job-level elastic scaling
% is presented in \cref{sec:design_loaning} and \cref{sec:design_elastic},
% respectively.

\noindent\textbf{Overall architecture.} \sys is a GPU cluster scheduler that
exploits capacity loaning with elastic job scheduling.
% the elasticity of cluster resources and training jobs.
% works across both training and inference clusters.
It runs on top of a cluster resource manager such as
YARN~\cite{vavilapalli2013apache} and Kubernetes~\cite{Kubernetes:Website} to
execute its decisions.
\begin{figure}[t]
    \centering
    \includegraphics[width=\linewidth]{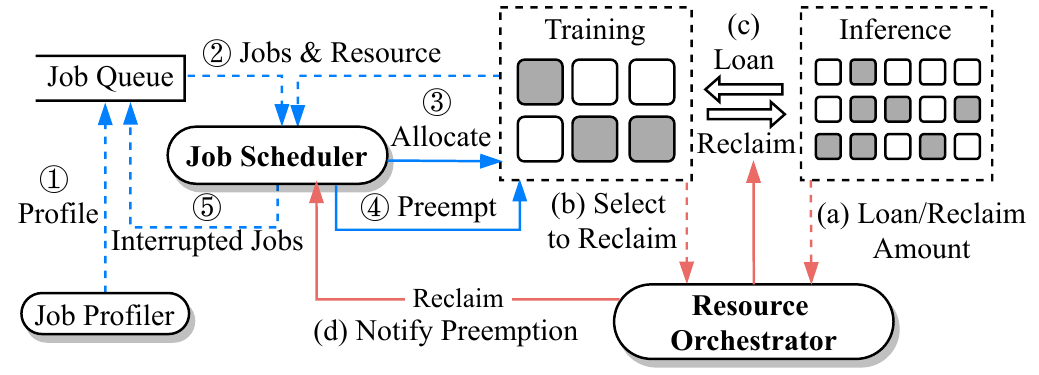}
    \vspace{-6mm}
    \caption{\sys system architecture. Solid lines indicate control flow and
    dashed ones data flow. Red lines represent capacity loaning workflow, while
    blue ones elastic scaling workflow. Each square represents a GPU server; the
    gray ones are in use.}
    \vspace{-6mm}
    \label{fig:system_architecture}
\end{figure}
\begin{comment}
We presume that the inference cluster scheduler dynamically estimates the
capacity needed to meet the latency, GPU utilization
\cite{KubernetesAutoscale:Website}, or other performance targets, based on the
predicted inference traffic \cite{258862, 10.1145/3341301.3359658,
crankshaw2017clipper}.
It then determines the amount of resources (1) that are available for loaning,
or (2) that needs to be reclaimed from training if any.
This way the inference performance is not affected by capacity loaning.
\end{comment}

Figure~\ref{fig:system_architecture} presents \sys's architecture.
At the cluster level, the \textit{resource orchestrator} obtains instructions
from the inference cluster about the number of servers to loan or reclaim (a),
determines which servers shall be returned for reclaiming (b), and
commands the underlying resource manager to move the selected servers
virtually across management boundaries to enforce the decisions (c).
When the orchestrator reclaims on-loan servers, it may need to preempt the
training jobs running on them (d). Job preemption is executed via the job
scheduler.

At the job level, jobs are submitted to the job queues.
The job profiler estimates the workload \circled{1} after
jobs are enqueued.
The \textit{job scheduler} \circled{2} {periodically} collects job status
and resource usage of the training cluster.
Then it \circled{3} computes the resource allocation and placement
decisions for each job.
Meanwhile, it gets preemption instructions from the
orchestrator, interrupts the running jobs \circled{4}, and puts them back to
the job queues \circled{5}.
Job launching, scaling and interruption actions are again executed by the resource manager.
Job scheduler works periodically in a much smaller interval than the
orchestrator in order to better handle job dynamics.

% \ljm{Modification: add system setup title, mention limited elasticity}
%
% \noindent\textbf{System constraints.}
Since \sys mainly deals with the training cluster and does not interfere with
inference cluster scheduling, we use ``jobs'' to simply refer to training jobs
hereafter without ambiguity.
The basic unit of capacity loaning is a physical server.
This is to prevent training jobs from interfering with the inference jobs on the same server~\cite{gilman2021characterizing}.
% malfunctioning training job may interfere with the customer-facing
% inference job on the same server.

\noindent\textbf{Key questions.}
\sys's design is centered around two key questions.
\begin{itemize}[leftmargin=*]
    % \item
    % How much capacity should we borrow from the inference cluster? The
    % amount of on-loan resources should be minimized to reduce the moving
    % overheads.
    \item \textbf{Server reclaiming.}
    Which servers should be returned so that the number of preempted
    jobs is minimized, when some on-loan servers need to be reclaimed?
    \item \textbf{Job scheduling.}
    How should we determine resource allocation across jobs, and how do we place a job's workers on servers, when some jobs are elastic and some servers are loaned from the inference cluster?
    % How should we schedule jobs to minimize the
    % overall JCT considering that cluster resources are dynamic, and some jobs
    % are elastic? \ljm{How elastic scaling helps in server reclaiming?}
\end{itemize}
% We believe neither question has been studied in previous work.
We now present how we address them with \sys's detailed design in
\cref{sec:design_reclaim} and \cref{sec:design_elastic},
respectively.

%% file: loaning.tex
\section{Loaning and Reclaiming Inference Servers}
\label{sec:design_reclaim}

\sys moves resources dynamically across inference and training clusters to improve utilization and training performance.

% \ljm{Move traffic estimation here and Briefly talks about resource loaning}
We presume that the inference cluster scheduler dynamically estimates the
capacity needed to meet the latency, GPU utilization
\cite{KubernetesAutoscale:Website}, or other performance targets, based on the
predicted inference traffic \cite{258862, 10.1145/3341301.3359658,
crankshaw2017clipper}. Inference workloads are able to grow or shrink their
containers along with the incoming traffic. Inference scheduler informs \sys's
resource orchestrator of (1) the amount of resources available for loaning when traffic is low, and (2) the amount of resources to be reclaimed from training in busy hours if any.
That is, the inference cluster scheduler completely determines when and which servers to lend, and when and how many servers to ask back, based on its own policy.
This way the inference performance is not affected by capacity loaning.

The key question for the training scheduler is the reclaiming mechanism as
mentioned in \cref{sec:design}, i.e. which on-loan servers should be returned
given the number of servers needed by the inference scheduler. This matters
because reclaiming a server entails preempting all its running jobs immediately. 
A job with checkpointing incurs overheads to save and load the checkpoint
when resuming training later. 
If the job does not involve checkpointing
\cite{264850}, its entire progress is lost and training has to restart from the very beginning. 
To use checkpointing or not is solely controlled by the user.
Therefore, the scheduler needs to minimize preemptions by strategically picking the servers to return, especially considering that it is actually common in our environment for jobs to not have checkpointing.
\noindent\textbf{Minimizing preemptions. }
% \noindent\textbf{Minimizing preemption when inevitable.}
Vacating an on-loan server means its jobs are preempted in a cluster with no elastic jobs.
We start with how \sys minimizes inevitable preemption under this case.
and will explain how elastic scaling plays its part in minimizing preemptions in \cref{sec:placement_design}.

% In case the flexible group alone is insufficient, preemption becomes
% inevitable: vacating any remaining on-loan servers means its jobs are preempted
% for certain.
Denote the number of servers that need to be returned at this point as
$N_R$.
Our problem is to pick $N_R$ on-loan servers---which host
inelastic jobs' workers---in order to minimize
preemptions.
More concretely, we choose to minimize the number of preempted jobs so fewer
users are affected.
This implies when reclaiming a server, we prefer the one with a big job to the
one with a few small ones.

The problem closely resembles the classic knapsack problem (i.e. the 0-1
knapsack problem):
The number of servers to reclaim $N_R$ can be considered as the capacity of the
knapsack; each server consumes one unit capacity, and the number
of running jobs is each server's preemption cost (i.e. value).
However, the server's preemption cost actually has inter-dependencies that make the
problem more difficult.

\begin{figure}[t]
\footnotesize
    \centering
    \includegraphics[width=0.9\linewidth]{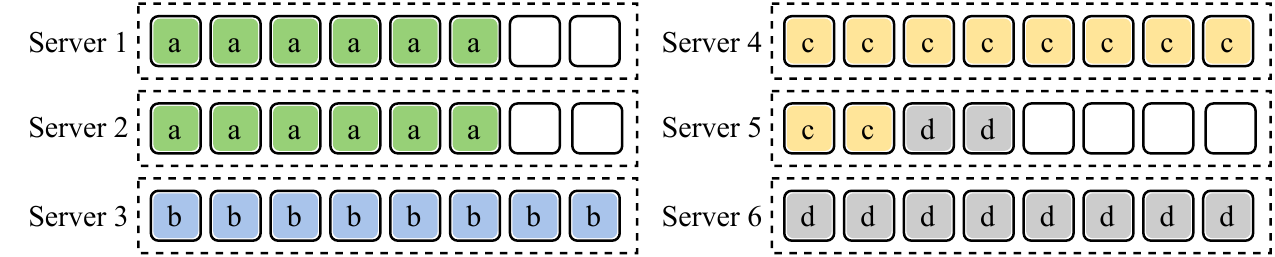}
    \vspace{-2mm}
    \caption{A reclaiming example. Each server has 8 GPUs.
    GPUs in-use are indicated by the job ID inside each square.
	GPUs with the same color are hosting the same job.
    }
    \vspace{-2mm}
    \label{fig:reclaim_example}
\end{figure}

Consider an example as depicted in Figure~\ref{fig:reclaim_example}.
Table~\ref{table:server_cost} shows each server's preemption cost as the number of its
running jobs (second column).
Suppose we need to reclaim two servers.
Servers 1 and 2 are obviously the optimal choice with one preemption.
Yet the corresponding knapsack problem would select any two 1-cost servers such
as 3 and 4 which lead to more preemptions.
The issue here is that in our problem the costs of servers are coupled when they
host the same job(s), whereas in the 0-1 knapsack problem the cost is
independent of each other.
Reclaiming server 1 for instance results in an idle server 2 whose cost
becomes 0 instead of 1.
% Thus when a server is reclaimed the costs of related servers need to change

\begin{table}[t]
\centering
\small
% \vspace{-2mm}
\resizebox{0.95\columnwidth}{!}{
     \begin{tabular}{|c|c|c|c|}
		\hline
		Server & \# running jobs & sum of job's GPU fraction & sum of job's
		server fraction \\
		\hline
		1      & 1         & 0.5       & 0.5       \\ \hline
		2      & 1         & 0.5       & 0.5       \\ \hline
		3      & 1         & 1         & 1         \\ \hline
		4      & 1         & 0.8       & 0.5       \\ \hline
		5      & 2         & 0.4       & 1         \\ \hline
		6      & 1         & 0.8       & 0.5       \\ \hline
		\end{tabular}
	}
  \vspace{-2mm}
  \caption{Different definitions of server preemption cost for the reclaiming example in
  Figure~\ref{fig:reclaim_example}.}
  \vspace{-6mm}
  \label{table:server_cost}
\end{table}
Knapsack problem with dependent item values is known to be NP-hard
\cite{mougouei2017integer}.
When $N_R$ is one server, selecting the one with fewest preemptions is simply
by iterating all the on-loan servers.
Given an $N_R$ larger than a single server,
we propose to resolve the dependency by treating it
as part of the server preemption cost. One possible way is to define server
preemption cost as the sum
of the GPU fractions of each job on the server. For instance, server 4's cost
would be 0.8 as it hosts 80\% of job c's GPUs, and server 5's cost is 0.4
(0.2+0.2) as shown in Table~\ref{table:server_cost}.
One can immediately see
that this does not work well as it does not capture the job count. It causes
server 5 to be selected with the least cost, which actually leads to two
preemptions. Thus we choose to define server preemption cost as the sum of the
server fractions of each job as shown also in Table~\ref{table:server_cost}.
This way server 5's cost is 0.5+0.5=1, i.e. the highest.

Once the preemption cost of each server is computed, the orchestrator selects
the servers using the following heuristic:
it iteratively picks the server with the lowest preemption cost, preempts its jobs by
removing them from all their servers, and updates the cost of these servers
correspondingly, until $N_R$ servers are vacated.
% Algorithm~\ref{algo:reclaim} in
\Cref{app:reclaim_select}  summarizes the complete reclaiming design.

%% file: elastic.tex
\section{Job Scheduling}
\label{sec:design_elastic}
\sys schedules jobs---both inelastic and elastic---to reduce
overall JCT by utilizing resources as efficiently as possible.
We start by explaining the challenge due to elasticity (\cref{sec:scheduling_challenge}).
Then we present the solutions to the two facets of our scheduling problem.
%Our scheduling problem has two facets.
The first is \textit{resource allocation} with both elastic and inelastic jobs,
i.e. how to determine the number of workers each job gets (\cref{sec:greedy_allocation});
the second is \textit{placement}, i.e. which servers to place a job's workers on (\cref{sec:placement_design}).
%The primary challenge is elastic jobs' resource allocation.
% In a scheduling operation, two components are involved, job
% scheduler and job profiler.
Here we assume that training throughput scales linearly with the
number of workers within the scaling range, as discussed in~\cref{sec:elastic}. 
% i.e. job's running time is inversely
% proportional to resources allocated, 
% \hx{what if you can model it as a non-linear function?} \ljm{mentioned in discussion}

\subsection{Challenge of Elasticity}
\label{sec:scheduling_challenge}

Job elasticity presents a unique challenge to resource allocation.
Conventional schedulers either deal with jobs with fixed demands,
or ones that can arbitrarily scale \cite{265013, qiao2020pollux}.
However, for jobs with {\em limited elasticity}~\cite{or2020resource},
the question of how to arbitrate resources so as to minimize average JCT is intricate.

% \ljm{Discussion: whether this is still required as existing work briefly
% discuss the sub-optimality of classical algorithms (but without range limits)}
Let us consider a simple example as shown in Table~\ref{table:elastic_example}. There are two elastic jobs with different minimum
running times when allocated their maximum demand.
Assume the cluster has eight workers in total.
Table~\ref{table:elastic_example_result} shows three common allocation
strategies and the corresponding JCT performance.
% Note we only show the initial allocation; when the first job finishes the
% other is immediately allocated the remaining resources as much as possible.
In solution 1, we favor job A by giving it the maximum demand; in solution 2, we
favor B instead; and in solution 3 we equally allocate resources to them.
All three strategies lead to different JCTs and the difference between
the worst and best is 24\%, demonstrating that inefficient allocation
can lead to poor JCT performance.

\begin{table}[t]
  \centering
  % \vspace{-3mm}
  \resizebox{0.5\columnwidth}{!}{
        \begin{tabular}{|c|c|c|c|c|}
        \hline
        % Job & $w_{min}$ & $w_{max}$ & Min. running time \\ \hline\hline
        % A & 2 & 10 & 2.4  \\ \hline
        % B & 2 & 8 & 2.5 \\ \hline
        Job & $w^{min}$ & $w^{max}$ & Min. running time \\
        \hline\hline
        A & 2 & 6 & 50  \\ \hline
        B & 2 & 6 & 20 \\ \hline
        \end{tabular}
  }
  \vspace{-2mm}
  \caption{Two elastic jobs and their demand information. Jobs completes in
  min. running time when allocated with $w^{max}$ workers.}
  \vspace{-3mm}
  \label{table:elastic_example}
\end{table}

\begin{table}[t]
\centering
% \vspace{-2mm}
\resizebox{0.8\columnwidth}{!}{
    % \begin{tabular}{|c|c|c|c|c|c|}
    \begin{tabular}{| c | *{6}{>{\centering\arraybackslash}p{1.3cm} |}}
    \hline
     \multirow{2}{*}{Solution} & \multicolumn{2}{c|}{Initial allocation}
     & \multicolumn{2}{c|}{JCT}
     & \multirow{2}{*}{\shortstack[c]{Average \\ JCT}} \\ \cline{2-5}
       & A   & B & A    & B     &  \\ \hline
      1 & 6  & 2 & 50 & 53.33    & 51.67 \\ \hline
      2 & 2  & 6 & 63.33   & 20  & 41.67 \\ \hline
      3 & 4  & 4 & 60 & 30    & 45 \\ \hline
    \end{tabular}
  }
  \vspace{-2mm}
    \caption{Possible resource allocation results for the two jobs when they
    share a cluster that can host 8 workers. Only the initial allocation
    is shown; once the first job finishes, the other is immediately
    allocated more
    resources as much as possible. Three solutions lead to very different
    JCTs. }
    \vspace{-4mm}
    \label{table:elastic_example_result}
\end{table}

\begin{table}[t]
    \centering
    % \vspace{-3mm}
    \resizebox{0.95\columnwidth}{!}{
          \begin{tabular}{|c|c|c|c|c|c|c|}
          \hline
          Job & $w^{min}$ & $w^{max}$ & Min. running time &
          JCT when favored & Avg. JCT \\
          \hline\hline
          A & 2 & 3 & 100 & A: 100, B: 24 & 62\\ \hline
          B & 2 & 6 & 20 & A: 106.67, B: 20 & 63.33 \\ \hline
          \end{tabular}
    }
    \vspace{-2mm}
    \caption{A counter example with two elastic jobs, where prioritizing A
    with longer running time is actually better for JCT.}
    \label{table:elastic_counter_example}
    \vspace{-7mm}
\end{table}
\noindent\textbf{Classic algorithms are not optimal. }
One may be wondering if
the classic shortest (or smallest) job first strategies would work here.
At least in the example of Table~\ref{table:elastic_example}, the optimal
allocation is indeed to first satisfy job B, which has the shortest running
time. Yet, we can construct a counter example as depicted in Table~\ref{table:elastic_counter_example} to show that this does not always
work.
We slightly modify job A to have a maximum demand of 3, and minimum
running time of 100; other setup is identical to
Table~\ref{table:elastic_example}. In this case, if we satisfy B first, the
average JCT (63.33) is actually worse than satisfying A's demand first (62).

Intuitively, shortest job first, or SJF, is designed for fixed job running
times with the intuition that each job should be given the least queuing time,
which is the only variable in computing JCT \cite{osconcept}. 
% Therefore, SJF schedules each job
% in a binary operation. 
In our case, job running time itself varies along with the
resource allocated, which in turn affects the overall JCT and makes the problem more complex. 

More specifically, the above examples reveal two characteristics of elastic
job's running time that SJF cannot handle. 
(1) Elastic scaling complicates the job sorting decision of SJF. 
% In the second case of Table~\ref{table:elastic_counter_example}, 
Since job running time varies with resource allocated, it is no longer apparent that we simply sort them based on their minimum running time. 
As shown already, doing so does not lead to the optimal result. 
% SJF still assumes fixed running time and sorts jobs with their minimum running time.  
% When jobs are elastic, it is questionable to naively sort with minimum running time . 
% (1) Bounded elasticity restrains the
% reduction of running time. In the first case where job A has a maximum demand of
% 6, the job completion order is consistent with the preference in the initial
% allocation. That is, whenever a job gets its maximum demand, it completes ahead
% of the other. Yet in the latter case
% (Table~\ref{table:elastic_counter_example}), A still completes after B though A
% is favored in the initial allocation, because its maximum demand of 3 limits the
% running time reduction it can achieve.
(2) The resource efficiency of each job is different. In Table~\ref{table:elastic_counter_example}, job A has a larger workload (i.e. product of maximum demand and minimum running time) than B, implying that the running time
improvement of A is larger than that of B if both are given the same number of
workers. Even though the resource
allocation difference is merely one when we prioritize different jobs, job A's
running time contributes to a 6.67-second JCT reduction while job B's only
increases by 4 seconds.

In the simplest two-job case, we can analyze the outcome of different
allocation strategies. 
In \Cref{app:elastic_theoretical_analysis}, we provide a complete theoretical analysis of this case.
Allocation in the general case is undoubtedly more complicated
with more elastic jobs plus inelastic jobs, as the optimal
strategy requires enumerating the exponentially many possible resource allocations. 
% plus the presence of inelastic jobs.
Our quest in the following is therefore to find a good heuristic for the
problem.

\begin{comment}
\ljm{Add challenge of dynamic resource pool}
\noindent\textbf{Dynamic resource pool.}
In \cref{sec:comparison}, we explain the uniqueness of \sys. \sys need
to schedule jobs in a dynamic resource pool. Existing DNN cluster
scheduler \ljm{TODO: add citation} generally allocate job resource
based on a certain metric regardless of the cluster capacity. Tiresias
orders jobs by attained service; AFS adopts job throughput gain; Pollux
computes both job throughput and efficiency. We observe that
a good resource allocation varies when the cluster capacity is changing.
\ljm{TODO: see if an example can be constructed}
Hence \sys seeks to adopt a good heuristic considering the
changing cluster capacity.
\end{comment}

\subsection{Two-Phase Resource Allocation}
\label{sec:greedy_allocation}

\noindent\textbf{Intuition: Prioritize inelastic workload.}
To ease the challenge of elasticity, our insight is that an elastic job has two
types of demand: a \textit{base demand} that is inelastic in nature, i.e. the
minimum demand, and a \textit{flexible demand} that is elastic.
They should be treated separately:
The base demand essentially corresponds to an
inelastic job whose allocation strategy is binary, and not allocating resources
to it incurs more queuing delay to the job.
In contrast, the flexible demand can be unfulfilled without serious impact
since the job is still making progress with base demand.

Therefore, we treat the \textit{inelastic} workload, including elastic jobs' base demands and inelastic jobs, as the first class citizen.
We schedule them first with all available resources to minimize the average JCT.
This also avoids starvation.
Then in phase two, we consider the flexible demand of elastic jobs to
fully utilize the remaining resources from phase one.
%This intuition is also applied in solving the placement sub-problem as will be
%shown in \cref{sec:placement_design}.

% One of the key difficulties is that jobs can be scaled whenever the cluster
% available resource changes.
% Those dynamics cannot be formulated in a single scheduling operation.
\noindent\textbf{Setup and assumptions. }
% When designing the heuristic,
We focus on solving the offline setting myopically where the set of jobs and
resources are given, and cope with the job dynamics and cluster capacity change by periodically performing scheduling in high frequency.
This is common in the literature \cite{fuxi, feitelson1998metrics}.
Our scheduling solution is non-preemptive to
minimize disruptions to training; preemption only happens during reclaiming when
it becomes inevitable as in \cref{sec:design_reclaim}.
Thus at a scheduling epoch, the set of available resources refer to idle
GPUs and GPUs being used by flexible workers for resizing (including on-loan GPUs), and the set of jobs include those waiting in the queue and running elastic jobs (only flexible workers).
The on-loan inference GPUs are normalized relative to training GPUs when calculating the resource capacity.
%We consider GPU only which is the bottleneck for most jobs.
% We normalize the total on-loan inference GPU capacity in the allocation problem,
% and leave the placement problem to \cref{sec:placement_design}.

% \ljm{%(Solution2, predicted running time)
We rely on job's running time information (min. running time for
elastic jobs), which can be predicted with profiling and ML methods~\cite{habitat, 254479}.
% \hx{add something about the prediction}.
% We use two separate models to estimate the training time of routine recurring
% jobs and exploratory ad-hoc jobs.
Details can be found in \cref{sec:implementation} and the effect of prediction
error will be discussed in \cref{sec:testbed_results}.
% }

% \ljm{Our job profiler profiles the job running time mainly
% by two methods, retrieving similar jobs in the history and running a prediction
% model. We usually estimate the running time of routine workloads from history
% records and predict for explorative ad-hoc workloads.}

% Our solution
% propose finding the local optimal solution in a single scheduling and allowing
% future scheduling to adjust the allocation with the job status.
% This frees job scheduler from
% searching for the optimal solution while considerinng a constantly-changing cluster.

% the thought to study the theoretical case
% Our analysis in \cref{sec:scheduling_challenge} reveals that job elasticity
% brings significant challenges here.

\noindent\textbf{Two-phase heuristic design. }
We now elaborate our heuristic.
The problem in phase one is how to minimize average JCT for jobs
with fixed demands and known running times, for which we adopt the shortest job
first (SJF) policy \cite{eilon1977minimising} which is a sensible and commonly
used solution. As long as there are idle GPUs and pending jobs, we schedule job
$j^*$ with the smallest running time. If demand of $j^*$ exceeds the remaining
capacity, we remove it from the pool and continue.

\begin{figure}[t]
    \centering
    \includegraphics[width=0.35\columnwidth]{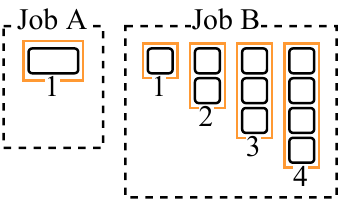}
    % \captionof{figure}{Items are represented by rectangles with numeric index. Each item
    % contains some worker(s) (circle). Items are grouped by jobs. During selection,
    % at most one item can be selected from a group.}
    \qquad
    \resizebox{0.5\columnwidth}{!}{\begin{tabular}[b]{|c|c|c|>{\centering\arraybackslash}p{3cm}|}
    \hline
    Group & Item & Weight & JCT Reduction Value \\ \hline\hline
    A& 1 & 2& 50 \\ \hline
    \multirow{4}{*}{B} & 1 & 1 &  20\\ \cline{2-4}
    & 2 & 2 &  30\\ \cline{2-4}
    & 3 & 3 &  36\\ \cline{2-4}
    & 4 & 4 &  40\\ \hline
    \end{tabular}}
    % \captionsetup{labelformat=andtable}
    \vspace{-2mm}
    \caption{Item weights and JCT reduction values for jobs in
    Table~\ref{table:elastic_counter_example}.
    Here, we assume job A needs 2 GPU per worker and job B 1 GPU per worker.}
    \label{fig:elastic_pack_example}
    \vspace{-6mm}
\end{figure}

% explain the input later. available resources is C - running_inelastic; jobs
% are pending jobs + running elastic ones
Phase two is more interesting.
We must determine the number of additional GPUs elastic
jobs get to maximize the JCT reduction.
Note elastic jobs here include those already running.
It turns out this problem can be transformed into 
\textit{multiple-choice
knapsack problem} \cite{sinha1979multiple}:
The knapsack's capacity is the number of remaining GPUs.
An elastic job $j$ is a group with $w^{max}_j-w^{min}_j$ items, each representing a possible allocation result for $j$'s flexible demand.
An item's weight is the number of GPUs required for this allocation, and its
value is the JCT reduction it brings over the job's maximum running time.
Figure~\ref{fig:elastic_pack_example} illustrates this transformation with the
two-job example in Table~\ref{table:elastic_counter_example}.
The problem is to pack the items into the knapsack so that the total value is
maximized, with the constraint of taking exactly one or zero item from each
group.

The multiple-choice knapsack problem, similar to the classical knapsack, is
NP-hard and often solved by dynamic programming which runs in pseudo-polynomial
time \cite{sinha1979multiple}.
With a moderate number of GPUs and jobs, dynamic programming can usually solve
the instance efficiently.
% We find that the longest solution time in our evaluation is 0.02s with 354
% items and 245 GPUs which is much shorter than a typical job's
% time use. 
% Algorithm~\ref{algo:elastic} in
\Cref{app:scaling_allocation} shows the complete algorithm.

\subsection{Worker Placement}
\label{sec:placement_design}
Given the allocation results, i.e. number of workers each job gets,
we need to determine the placement of each worker. Our fundamental strategy is bin packing with best-fit decreasing heuristic\cite{panigrahy2011heuristics}. 
Elastic jobs are preferably placed on the inference servers to maximize the potential for scaling in during reclaiming and reduce job preemptions while inelastic jobs are placed on training servers 
whenever possible.
\Cref{subsec:simulation} presents how this placement strategy helps 
reduce the job preemptions in reclaiming the on-loan servers when there are elastic jobs.

% \begin{comment}
Given the allocation results, i.e. number of workers each job gets,
we still need to determine the placement of each worker to
complete scheduling.
Our goal is to reduce fragmentation.
The primary concern is the mix of inelastic and elastic jobs as well as the transient on-loan servers with different GPUs.

Our fundamental strategy is bin packing with best-fit decreasing (BFD)
heuristic\cite{panigrahy2011heuristics}.
% Meanwhile, we consider the locality of jobs and the choice of servers.
% We categorize the jobs into two types, small jobs whose demand is less than or equal to a single server's (i.e. small jobs)
% capacity and job demand greater than a single server's capacity (i.e. large jobs).
% \sys's job scheduler guarantees the locality of small jobs first.
% It makes the best effort to place a small job within a single server to avoid inter-machine communication.
Jobs are sorted in decreasing order of their per-worker GPU demand as
GPU is most likely the bottleneck resource for training.
Starting from the largest job, we place each worker of the job into a non-empty
server that best fits its demand; if none has sufficient remaining resources, we place it on a new server.
If the job is elastic, we prefer to place it on inference servers in order to maximize the potential for scaling in during reclaiming and reduce job preemptions.
If it is inelastic, we prefer to placing it on training servers.
When placing elastic jobs, we also place their base and flexible demands on separate groups of inference servers so that during reclaiming (\cref{sec:design_reclaim}), \sys can release the server group for flexible demands first without any preemption to see if this alone is sufficient.
% \end{comment}

%% file: implementation.tex
%!TEX root = main.tex
\section{Implementation}
\label{sec:implementation}

We have implemented a prototype of \sys with about 3500 lines of Python.
The prototype works with our existing YARN and Kubernetes deployment to move
servers across clusters virtually, manage worker containers for training, and
monitor the status of servers and workers.
The reclaiming and scheduling algorithms are implemented following
Algorithms~\ref{algo:reclaim} and \ref{algo:elastic} in \Cref{app:scaling_allocation,app:reclaim_select}.
% The orchestrator interfaces with our production inference scheduler to obtain
% its in
% Our prototype can be readily deployed to manage production jobs.

We highlight key details of the implementation as follows.
% and implemented a simulator with 1500 lines of Python. We build the job
% scheduler on top of YARN and resoruce orchestrator on top of
% YARN and Kubernetes.

\begin{comment}
\noindent\textbf{Inference traffic prediction.}
As mentioned earlier in
\cref{sec:howtoloan}, resource orchestrator relies on the traffic
prediction of inference scheduler.
% Accurate prediction could let
% orchestrator promptly return the on-loan resources for inference
% service.
We implement a LSTM model with window size of 10
and two hidden layers. The training model apply Adam optimizer
and use MSE to compute loss. We predict the resource usage of next 5
mintues and compare the average resource usage with the ground truth.
The average loss is 0.00048. Based on the prediction result, the amount
of resource to move is decided.
\end{comment}
\noindent\textbf{Interface for capacity loaning.}
% During loaning, resource orchestrator need to update the available
% resource of each cluster once the operation is decided.
We create a \textit{whitelist} API to facilitate capacity loaning operations.
Both \sys's scheduler and the inference scheduler maintain their own whitelist
of servers under their control.
\sys's orchestrator adds on-loan servers to job scheduler's whitelist during loaning and removes the selected servers during 
reclaiming after its scheduler confirms they no longer have running workers.

\noindent\textbf{Data locality and resource isolation. }
\sys performs capacity loaning only between clusters in the same datacenter to ensure the network bandwidth across servers is consistently high.
% Meanwhile, HDFS is used for cluster file system.
% Theoretically, accessing HDFS data from any server is the same.
Also, the basic unit of loaning is a physical server so co-location of
inference and training jobs is not possible, and no additional isolation mechanisms are needed.
% Virtually moving servers between
% clusters requires no delicate resource isolation within a single server.

\noindent\textbf{Enable elastic scaling. }
We enable elastic training with a few modifications
to the ML frameworks. We embed a controller process to
each elastic job that coordinates the worker join and departure.
Base demand guarantees the gang scheduling
of minimum requests and the flexible demand shortens the running
time whenever possible while preserving loss convergence.
Some recent work \cite{or2020resource,258957} developed more complete scaling solutions that our implementation could also utilize.

\noindent\textbf{Job running time estimation. }
\sys's job scheduler relies on the running time estimated by the job profiler.
We implement a simple profiler based on the properties of the training jobs.
For the planned routine jobs, the profiler estimates based
on the history runs of the same job.
For those ad-hoc exploratory jobs, we adopt the prediction model in \cite{habitat} for estimation.

\noindent\textbf{Heterogeneous GPU training. }
As discussed in \cref{sec:motivation}, some training jobs can run on
heterogeneous GPUs experimentally. When this feature is turned on, \sys's job scheduler places them lastly on the remaining resources.
The actual scheduling logic for these jobs remains the same 
% as Algorithm~\ref{algo:elastic}
,
except that if they are elastic, their base demands are placed on training servers,
and flexible demands on inference servers whenever possible.
% Otherwise, they are treated equally but have a wider range of available resources.

\begin{comment}
% \ljm{Pendings
\noindent\textbf{Avoid redundant scaling.}
\sys's job scheduler needs to scale in running elastic jobs to their
base demand, and then computes the new allocation result which may scale out
some of these jobs again as mentioned in \cref{sec:greedy_allocation}.
This is to fully exploit elastic jobs to optimize resource allocation.
Thus to avoid the redundant scaling operations, our scheduler
creates a copy of the current job placement, and performs scheduling on
the copy without scaling any jobs yet.
The scheduler only commits scaling operations that are necessary to transition
from the current placement to the newly computed one.
\end{comment}
% It is likely that some elastic jobs do not necessarily require
% scale in when comparing with the previous placement and ultimate scheduling
% decision. Therefore, our job scheduler creates a copy of the current
% job placement and resource status and models scaling annd scheduling
% on the copy instead of actually scaling the running jobs. This can
% largely avoid redundant scaling to training jobs.
% }

%% file: evaluation.tex
%!TEX root = main.tex 
\section{Evaluation}
\label{sec:evaluation}

We evaluate \sys using both large-scale simulations 
and testbed experiments with
traces from our production clusters. 
% \begin{comment}
The highlights of our findings are:
\begin{itemize}
	
	\item In large-scale simulations, \sys's benefit is more salient with 1.53x and 1.50x reductions on average queuing time and JCT, respectively.
   Capacity loaning has a factor of 1.39 and 1.33
	 reductions in average queuing time and JCT. Elastic scaling leads to
	 reductions of 1.35x and 1.38x in average queuing time and JCT.

  % \hx{
  \item Compared to state-of-the-art scheduler Pollux \cite{qiao2020pollux}, \sys's scheduling algorithm brings 1.35x average queuing time and 1.42x average JCT reductions when both consider tuning the training hyperparameters.
  % \item \sys's scheduling algorithm can be integrated with Pollux's job 
  % agent \hx{that adjusts training hyperparameters according to resource allocated}. This leads to 1.35x average queuing time and 1.42x average JCT reductions over Pollux.
  \sys's reclaiming algorithm performs comparably against the optimal solution with only 1--3ms running time.
  % }
  %  \hxq{queuing delay or queuing time, be consistent (resolved)}
  % \item \sys scheduling algorithm is robust to inaccurate job running time estimation and presents queuing time and JCT reduction of 1.76x and 1.38x even when 60\% predictions are wrong. Its reclaiming algorithm performs comparably against the optimal solution with only 1–3ms running time.
  \item In testbed, \sys improves average job queuing time by
  1.38x and average JCT by 1.22x over the baseline without loaning or scaling.
  Preemption only happens to $\sim$9\% of the jobs in reclaiming
	with an average 63-second overhead.
\end{itemize}
These benefits are achieved with only $\sim$5\% of the jobs being elastic as discussed in \cref{sec:elastic}. 
% \hxq{how to fix this?}
% \end{comment}

\subsection{Setup}
\label{sec:setup}
\noindent\textbf{Traces.}
We rely on a 15-day job trace from one of our production training clusters with
3,544 GPUs (443 8-GPU servers).
There are 50,390 training jobs, and job running time
range from minutes to days.
We also use a GPU utilization trace from the inference cluster for the same time
period.
Part of the traces have been shown in Figures~\ref{fig:inference_util_diurnal}
and
\ref{fig:training_queue_util} already.

\noindent\textbf{Simulator.} We built a discrete-event simulator for evaluating
\sys at scale using job traces from production.
It simulates the cluster scale, hardware configuration, and all job events including arrival, completion, scaling, and preemption.
Job's running time in the
simulator is derived from actual training time in the traces.
{For elastic jobs, we compute its actual training time based on the traces which is inversely proportional to its resource allocation as discussed in~\cref{sec:design_elastic}.
We also evaluate \sys when jobs have imperfect scalability in~\cref{subsec:simulation}. }

% {We verify the fidelity of our simulator.  
% We add 63-second job preemption overhead as measured from testbed experiments (see \cref{sec:testbed_results}). 
% The simulation results conform to those on testbed; the average error rate on average JCT is <5\% across all schemes. 
% % More specifically, simulation shows 6.2\% and 3.4\% difference in average and 95\%ile JCT, and 3.5\% and 4.4\% difference in average and 95\%ile queuing 
% % time compared with testbed results.
% The error mainly stems from the overhead of placing 
% workers and moving resources between clusters which the simulator 
% does not capture.} 
\begin{table*}[t]
  \centering
  \vspace{-2mm}
  \resizebox{15cm}{!}{
  \begin{tabular}{llllllllllll}
  \Xhline{1pt}
  \\[-1em]
  \multirow{2}{*}{\#} & \multirow{2}{*}{Scenario}         & \multirow{2}{*}{Solution} & \multicolumn{3}{c}{Queuing Time (s)}                                                 & \multicolumn{3}{c}{JCT (s)}                                                          & \multicolumn{2}{c}{GPU Usage}                              & \multicolumn{1}{c}{Preemption} \\ \\[-1em] \cline{4-12} \\[-1em]
                      &              											&                           & \multicolumn{1}{c}{Mean} & \multicolumn{1}{c}{Median} & \multicolumn{1}{c}{95\%ile} & \multicolumn{1}{c}{Mean} & \multicolumn{1}{c}{Median} & \multicolumn{1}{c}{95\%ile} & \multicolumn{1}{c}{Training} & \multicolumn{1}{c}{Overall \textsuperscript{1}} & \multicolumn{1}{c}{Ratio \textsuperscript{2}}      \\\\[-1em] \Xhline{0.7pt} \\[-1em]
  1	& Baseline  & --- 
  & 3072  & 55  & 8357  & 16610  & 791  & 82933  & 0.72  & 0.52  & 0        \\\\[-1em]\hline \\[-1em]
  2	& Basic     &      
  & 2008  & 25  & 3356  & 11089  & 567  & 56477  & 0.86  & 0.66  & 10.20\%  \\\\[-1em]
  3	& Advanced  & \sys                          
  & 1833  & 23  & 3238  & 10402  & 523  & 56553  & 0.87  & 0.69  & 7.05\%   \\\\[-1em]
  4	& Ideal     &                           
  & 1157  & 22  & 3197  & 8874   & 417  & 41146  & 0.92  & 0.72  & 5.57\%   \\\\[-1em]\hline \\[-1em]
  5 &           & Random                    
  & 2843  & 23  & 5471  & 14657  & 703  & 62912  & 0.76  & 0.64  & 20.89\%  \\\\[-1em]
  6 & Capacity Loaning & SCF                       
  & 2791  & 24  & 4991  & 14965  & 692  & 62451  & 0.76  & 0.66  & 18.74\%  \\\\[-1em]
  7 & & \tbg \sys                      
  & \tbg 2204  & \tbg 23  & \tbg 3418  & \tbg 12414  & \tbg 655  & \tbg 57982  & \tbg 0.76  & \tbg 0.66  & \tbg 12.34\%  \\\\[-1em]\hline \\[-1em]
  8 &\multirow{5}{*}{\shortstack[l]{Elastic Scaling \\ (Basic)}} & Gandiva                   
  & 3035  & 49  & 6632  & 15912  & 755  & 80567  & 0.79  & NA    & NA       \\\\[-1em]
  9 & & AFS                       
  & 2284  & 47  & 3488  & 15045  & 686  & 60883  & 0.95  & NA    & NA       \\\\[-1em]
  10 & & Pollux                    
  & 2791  & 58  & 5883  & 14534  & 721  & 72123  & 0.93  & NA    & NA       \\\\[-1em]
  11& & \tbg \sys                      
  & \tbg 2275  & \tbg 47  & \tbg 3475  & \tbg 12048  & \tbg 602  & 
  \tbg 57597  & \tbg 0.92  & \tbg NA    & \tbg NA       \\\\[-1em]
  12& & \tbg \sysP
  & \tbg 2054  & \tbg 43  & \tbg 2749  & \tbg 10229  & \tbg 564  & 
  \tbg 52458  & \tbg 0.91  & \tbg NA    & \tbg NA       \\
  \Xhline{1pt}
  
  \end{tabular}}
  \begin{tablenotes}
   \scriptsize
   \item (1) Overall GPU usage denotes the
    GPU utilization in both training and inference cluster. It is applied when the training cluster size is changing in capacity loaning.
   \item (2) Preemption ratio is the ratio between total number of preemptions and the total number of job submissions.
  \end{tablenotes}
  \vspace{-2mm}
    \caption{Simulation results using different solutions.}
    \vspace{-5mm}
    \label{table:simulator_stat}
  \end{table*}

  \noindent\textbf{Testbed.} Our testbed consists of four 8-GPU training servers
and
four 8-GPU inference servers. Each training server uses Nvidia V100 GPUs
with 32GB GPU memory and has 92 vCPU with 350 GB memory.
Each inference server uses Nvidia T4 GPUs with
16GB GPU memory and has 92 vCPU with 210 GB memory.
The resource management framework is YARN, and training data is stored
in HDFS.

\noindent\textbf{Scenarios.}
We consider various scenarios with different degrees of support for elastic scaling and heterogeneous training, both of which are not widely used today.
\begin{itemize}
	\item \textit{Basic}: Only large jobs with good scalability as discussed in \cref{sec:elastic} ($\sim$5\% of all jobs) support elastic scaling within a given range. No heterogeneous training. This corresponds to the status quo in our environment and is the default scenario.
	\item \textit{Advanced}: On top of \textit{Basic}, 10\% jobs can run on heterogeneous GPU with non-ideal performance.
	Specifically, heterogeneous training jobs only achieve at most 70\% of the ideal results. We present an empirical analysis of heterogeneous training performance in~\Cref{app:heterogeneous}.
	\item \textit{Ideal}: All jobs support scaling and heterogeneous training with ideal performance.
\end{itemize}

\noindent\textbf{Schemes compared.}
We compare \sys to the following schemes that represent the state-of-the-art
and/or the most common solutions to each sub-problem of \sys.
We consider two basic strategies for server reclaiming:
\begin{itemize}
  \item \textit{Random}: On-loan servers are randomly selected.
  %  to fulfill the request.
	\item \textit{Smallest (Job) Count First (SCF)}: The top-$k$ servers that host
  the smallest number of jobs are chosen.
\end{itemize}
We consider several solutions to elastic scheduling. Some are slightly modified to conform with our setup for elastic jobs. 
\begin{itemize}
  % \item \textit{Baseline}:
  % No capacity loaning or elastic scaling. Job scheduling uses FIFO.
  \item \textit{Gandiva}~\cite{xiao2018gandiva}:
  Elastic scaling is also mentioned in Gandiva. It exploits elasticity by scaling out jobs to utilize the remaining resources on servers whenever they are under-utilized.
  We consider under-utilization to be the period
	when there are available resources but no pending jobs.
  \item \textit{AFS}~\cite{265013}:
  It allocates one GPU to each job first and iteratively gives one more GPU to the job with the largest marginal throughput gain.
	We implement AFS by allocating base demand to each job first and allocating	one more worker to the job with the largest throughput gain per GPU.
	\item \textit{Pollux}~\cite{qiao2020pollux}:
  Pollux computes the goodput of training jobs and
	applies genetic algorithms to find the resource allocation. It also
	adjusts batch size to maximize goodput and learning rate based on Adascale\cite{johnson2019adascale}. We adopt the model distribution 
  listed by Pollux to capture the model goodput. 
\end{itemize}

\noindent We notice that Pollux's idea of tuning the hyperparameters according to allocated resources is orthogonal to job scheduling. To compare with Pollux fairly, we integrate this idea into \sys in~\Cref{sec:scaling_perf}:
\begin{itemize}
  \item \textit{\sysP}: Use \sys's job scheduler and adapt Pollux's job agent for job-level hyperparameter-tuning within the scaling range. Job agent adjusts model batch size and learning rate whenever job resource allocation changes. 
\end{itemize}

We consider \textit{Baseline} to be a FIFO cluster scheduler with no capacity loaning or elastic scaling.

% Finally, the following scheme captures how GPU clusters are usually managed
% today and serves as our baseline.
% \begin{itemize}[leftmargin=*]
  % \item \textit{Baseline}:
  % No capacity loaning or elastic scaling. Job scheduling uses FIFO.
% \end{itemize}

  \begin{table}[t]
    
    	\centering
    	\resizebox{0.7\columnwidth}{!}{
    	\begin{tabular}{@{}llll@{}}
    	\toprule
    	Scenario & Avg. Queuing Time & Avg. JCT & Preemption Ratio \\ \midrule
    	Basic    & 2231              & 13864    & 12.56\%          \\
    	Advanced & 1950              & 12237    & 10.28\%          \\
    	Ideal    & 1257              & 9751    & 11.89\%          \\ \bottomrule
    	\end{tabular}}
      \vspace{-2mm}
      \captionof{table}{Performance without special placement of elastic jobs. \sys naively places jobs based on the BFD heuristics.
      }
    	\label{table:noplacement_overall}
    	\vspace{-5mm}
  \end{table}
  \begin{figure*}
    \begin{minipage}[t]{\textwidth}
      \begin{minipage}[b]{0.34\textwidth}
        \vspace{-8mm}
          \resizebox{\columnwidth}{!}{
          \begin{tabular}{@{}lllllll@{}}
          \toprule
                   & \multicolumn{3}{c}{Queuing Time (s)} & \multicolumn{3}{c}{JCT (s)} \\ \cmidrule(l){2-7}
                   & Mean      & Median     & 95\%ile     & Mean   & Median  & 95\%ile \\ \midrule
          Baseline & 4573      & 1283       & 23351       & 11547  & 2122    & 60170   \\
          \sys     & 1029      & 272        & 7249        & 6832   & 1256    & 35604   \\ \bottomrule
          \end{tabular}}
          \vspace{-2mm}
          \captionof{table}{Queuing time and JCT of jobs running on on-loan servers.}
          \label{table:loan_jct}
      \end{minipage}%
      \hfill % maximize the horizontal separation
      \begin{minipage}[b]{0.32\textwidth}
        \includegraphics[width=\linewidth]
        {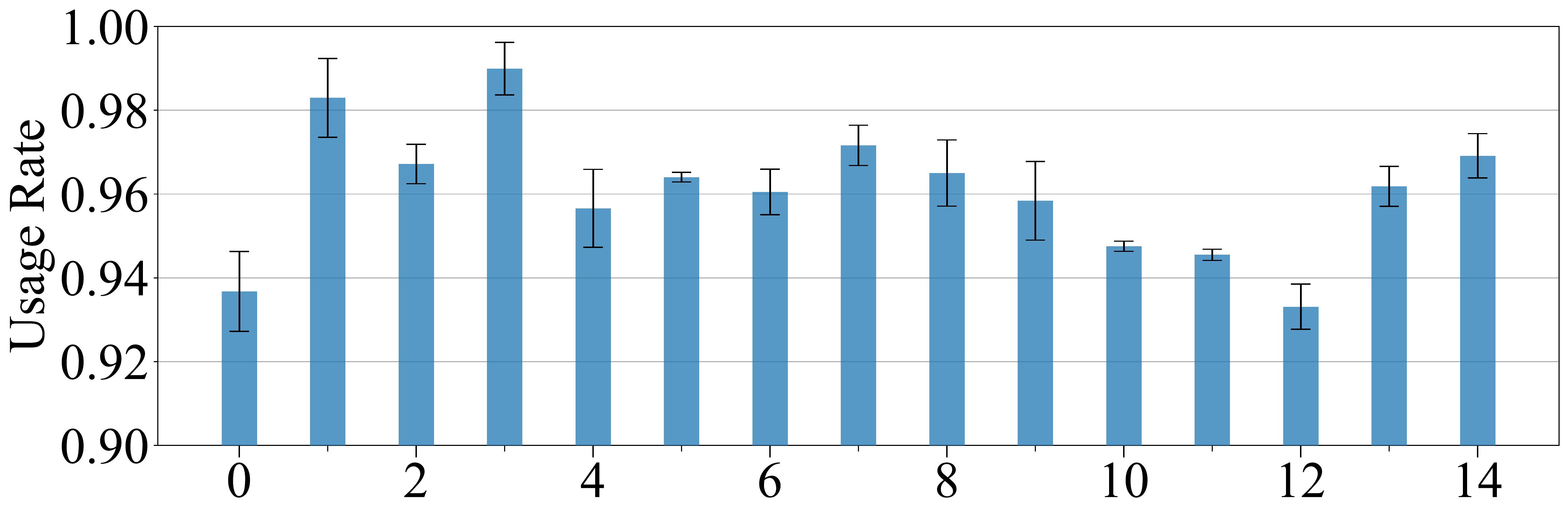}
        \vspace{-7mm}
        \captionof{figure}{The daily average resource usage of on-loan servers (monitored every 5 minutes). }
        \label{fig:reclaim_usage}
      \end{minipage}%
      \hfill
      \begin{minipage}[b]{0.30\textwidth}
        \includegraphics[width=\linewidth]
        {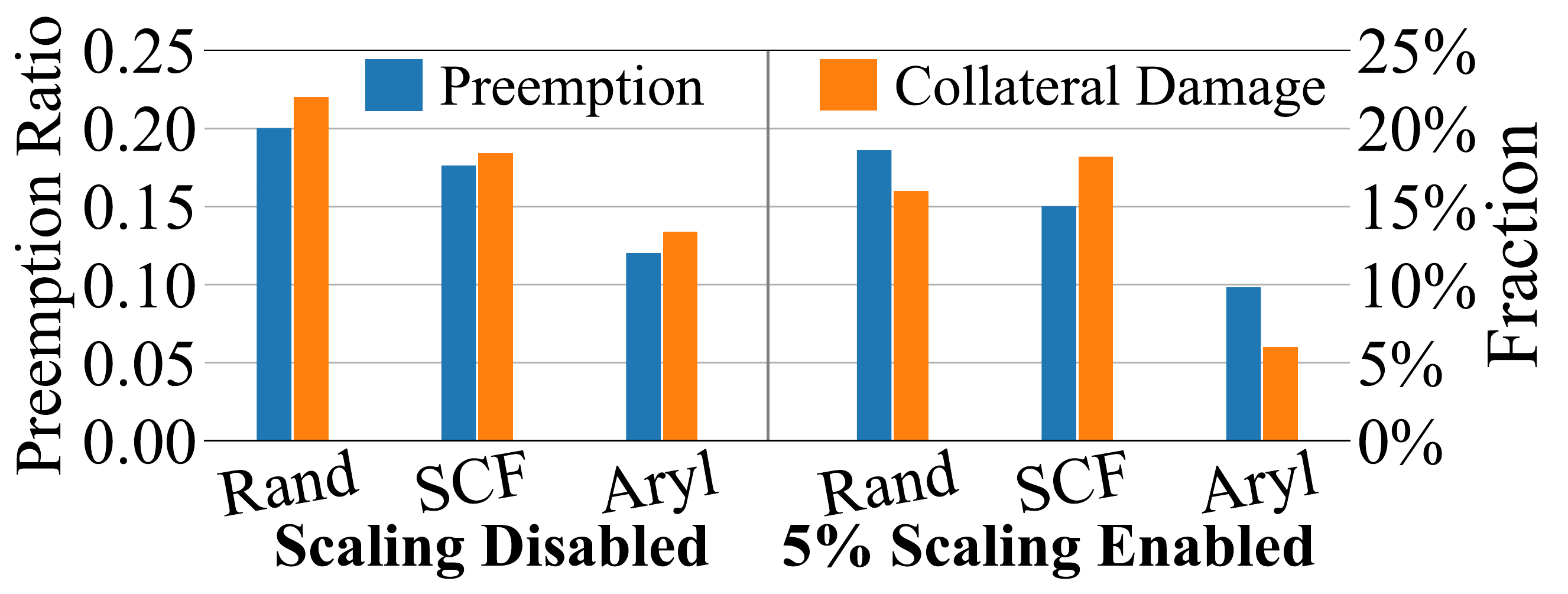}
        \vspace{-7mm}
        \captionof{figure}{Preemption ratio and average collateral
        damage comparison in simulator.}
        \label{fig:simulator_preemption}
      \end{minipage}%
    \end{minipage}
    \vspace{-8mm}
  \end{figure*}
\subsection{Overall Performance in Simulation}
\label{subsec:simulation}
We evaluate \sys thoroughly with large-scale simulation.
We first provide overall performance of \sys.
Analyses of its individual components are presented in
\cref{sec:loaning_perf} and \cref{sec:scaling_perf}.
% \ljm{Fidelity is moved to 7.1 under simulator.}
% \begin{comment}

\noindent\textbf{Simulator fidelity.}
To first establish its fidelity, we evaluate our simulator against the prototype
system in testbed with the small trace.
We add 63-second overhead whenever a job is preempted in simulation.
The simulation results are similar to testbed results, with a difference of
6.2\% and 3.4\% in average and 95\%ile JCT, and 3.5\% and 4.4\% in average
and 95\%ile queuing time.
The small difference mainly stems from the overhead of placing workers and
moving resources between clusters which the simulator does not capture.

\noindent\textbf{Cluster scale and workload. }
We use the full 15-day trace and the same cluster configuration as our production clusters.
% \end{comment}

\begin{figure}[t] 
  \begin{minipage}{0.3\textwidth}
  \centering
  \includegraphics[width=\linewidth]
  {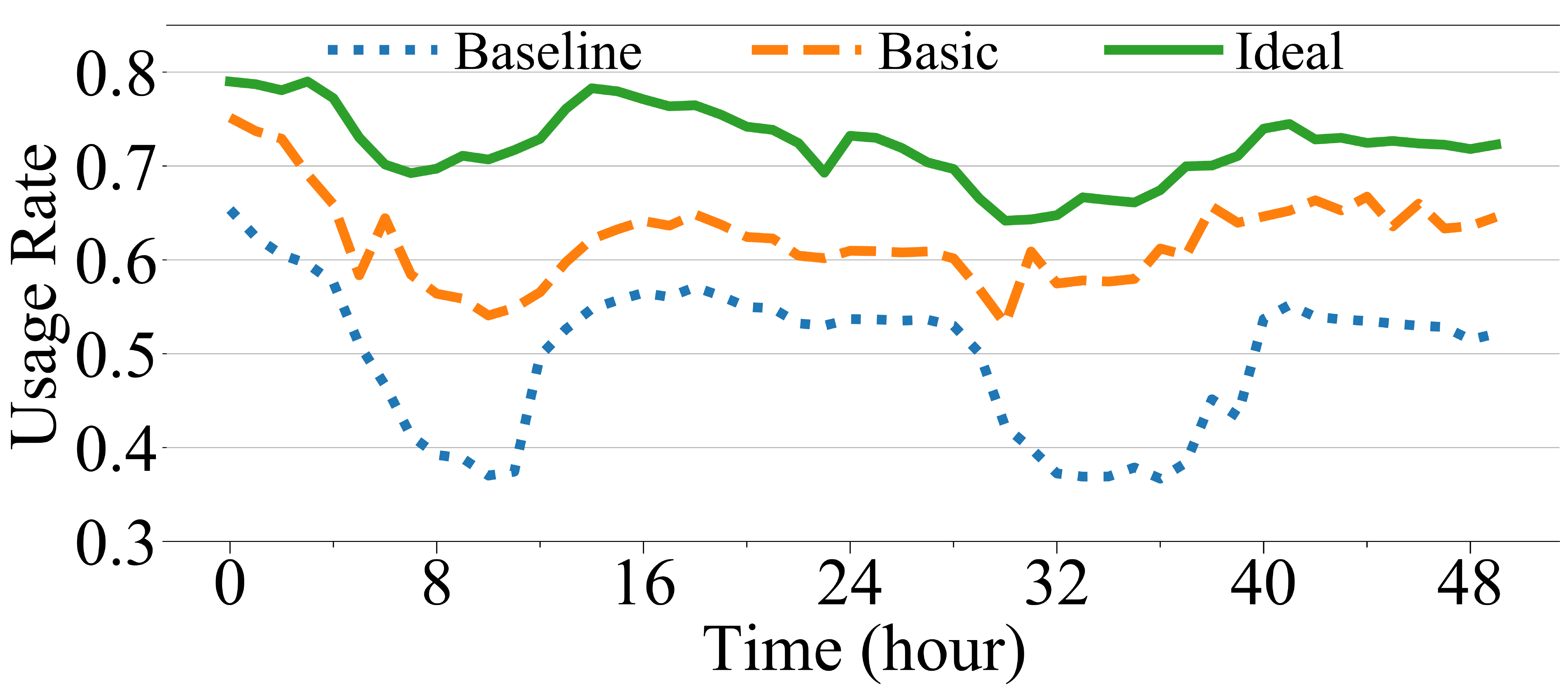}
  \vspace{-8mm}
  \captionof{figure}{Overall resource usage rate in different scenarios. }
  \label{fig:usage_rate}
\end{minipage}%
\hfill
\begin{minipage}{0.16\textwidth}
  \centering
  \vspace{-8mm}
  \includegraphics[width=\linewidth]
  {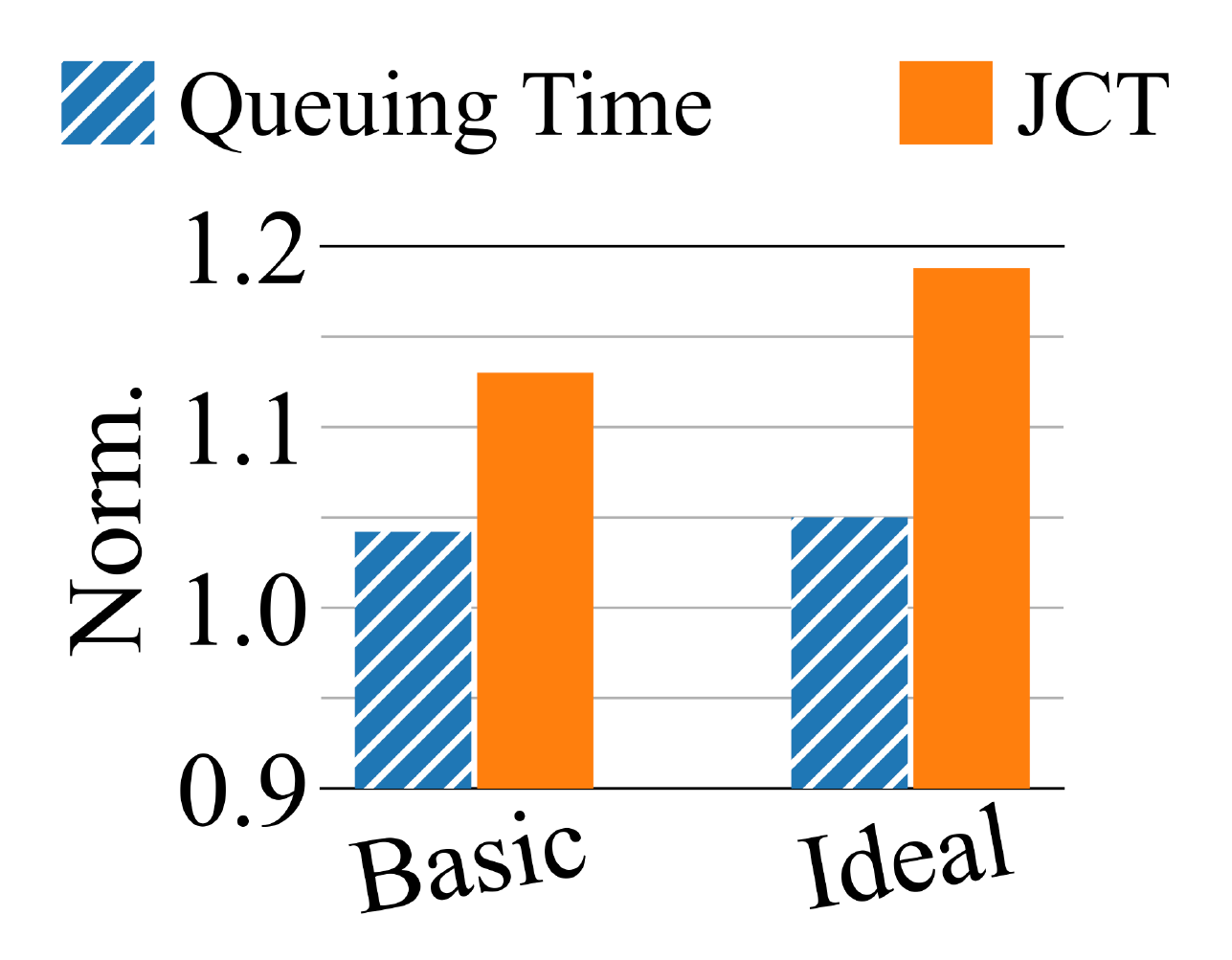}
  \vspace{-8mm}
  \captionof{figure}{Queuing time and JCT degradation 
  in \\ imperfect scaling. }
  \vspace{-8mm}
  \label{fig:nonlinear_perf}
\end{minipage}
\vspace{-4mm}
\end{figure}
\noindent\textbf{Queuing time, JCT, and cluster usage.}
Table~\ref{table:simulator_stat}
records the performance of \sys in different scenarios.
% \ljm{
Overall, queuing time and JCT are improved by 1.53x and 1.50x when comparing
to Baseline in the Basic scenario (row 2). 
The overall cluster usage is improved by 26.9\%.
In the {Advanced} case with non-ideal heterogeneous training, queuing time and JCT are reduced by 1.68x and 1.60x over Baseline and by 1.10x and 1.07x over \sys itself in the Basic scenario.
In the Ideal case which represents the performance upper bound, the average combined usage of the inference and training clusters is improved by 38.5\% (to 72\%) in Baseline.
Compared with the Basic case, average queuing time and JCT in the Ideal case show additional 1.12x and 37\% improvements by virtue of complete job flexibility and perfect performance scalability.
% }

Since the training cluster resource is dynamically changing, we depict the
hourly combined cluster usage for 48 hours in Figure~\ref{fig:usage_rate}.
The Baseline usage curve shows a clear diurnal pattern mostly attributable
to the inference cluster.
When capacity loaning is enabled, \sys improves the usage and flattens the curve; the most significant improvement is a 14\% usage increase 
between Basic and Baseline.
The combined usage does not reach 100\% as the inference cluster needs some headroom to gracefully handle the latency SLA.

\noindent\textbf{Gain from capacity loaning.}
We disable elastic scaling in \sys and evaluate its gain over
Baseline to understand the benefit of capacity loaning.
Table~\ref{table:simulator_stat} (row 7) shows that loaning alone
reduces average queuing time and JCT by 1.39x and 1.34x  over Baseline.
Loaning also improves the combined cluster usage from 52\% to 66\%.
We observe that the JCT improvement is not as significant as elastic scaling
(row 11). This is mainly because (1) loaning depends on idle inference resources
and its gain is less stable, and (2) compared to scaling, loaning itself does
not affect job running time.

% \begin{comment}
\noindent\textbf{How scaling helps capacity loaning?}
We now seek to understand how our two key ideas interact and complement each.
Scaling helps capacity loaning, especially in reducing
preemptions in reclaiming the on-loan servers.
With elastic scaling disabled, Table~\ref{table:simulator_stat} shows that preemption as percentage
of running jobs increases from 10.20\% (row 2) to 12.34\% (row 7). % \hxq{seems small}
We also observe that on average the flexible server group (hosting flexible
workers only) alone satisfies 52.6\% of reclaiming demand each time.
With more aggressive flexibility (row 4), preemption is reduced to 5.57\%
and satisfy 82.4\% of reclaiming demand each time.

In~\cref{sec:placement_design}, we discussed how \sys places elastic and inelastic jobs with on-loan servers in the cluster. In Table~\ref{table:noplacement_overall}, 
% \hxq{this table is in appendix now!}
% \ljm{(Resolved) Accidentally wrap Table 6 with Figure and wrong referencing}
we compare the placement performance in different scenarios without special treatment to elastic jobs, i.e. instead of grouping their flexible demand and placing to on-loan servers as much as possible, the scheduler places them to training servers first just like inelastic jobs.
The most significant difference is in preemption ratio.
Without grouping the flexible demand, preemption ratio increases by up
to 113\% in Ideal (compared to Table~\ref{table:simulator_stat} row 4). Preemptions also incur degradation to job runtime; for example average queuing time and JCT in the Basic case increase by up to 10\% and 20\%.
% \hxq{consider moving this entirely to appendix, only leave a pointer here to indicate we've considered it.}
% \end{comment}

\noindent\textbf{Impact of imperfect scaling}
Thus far we have assumed linear scalability of elastic jobs based on our empirical analysis in \cref{sec:elastic}.
Here we also evaluate \sys's performance where elastic jobs scale imperfectly with throughput loss.
Specifically, whenever a job scales beyond the midpoint of its scaling range, its throughput suffers additional 10\% loss from the ideal performance each time it scales one step further.
Figure~\ref{fig:nonlinear_perf} presents the queuing time and JCT 
degradation compared with linear scalability in Basic and Ideal scenarios.
In Basic, average queuing time and JCT are 4\% and 12\% higher than those with linear scalability (Table~\ref{table:simulator_stat} row 2).
Since \sys prioritizes base demand, queuing time does not degrade much even in the Ideal case.
Yet average JCT is inflated by 19\% to 10,564 seconds (compared to Table~\ref{table:simulator_stat} row 4).
\begin{figure*}
	\centering
  \includegraphics[width=0.5\linewidth]{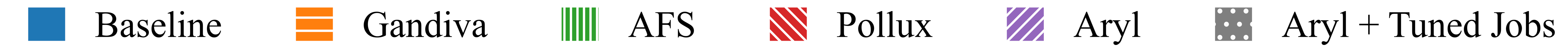}
	\vspace{-4mm}
\end{figure*}
\begin{figure*}
	\begin{minipage}[t]{0.245\textwidth}
	    \centering
	    \includegraphics[width=\linewidth]{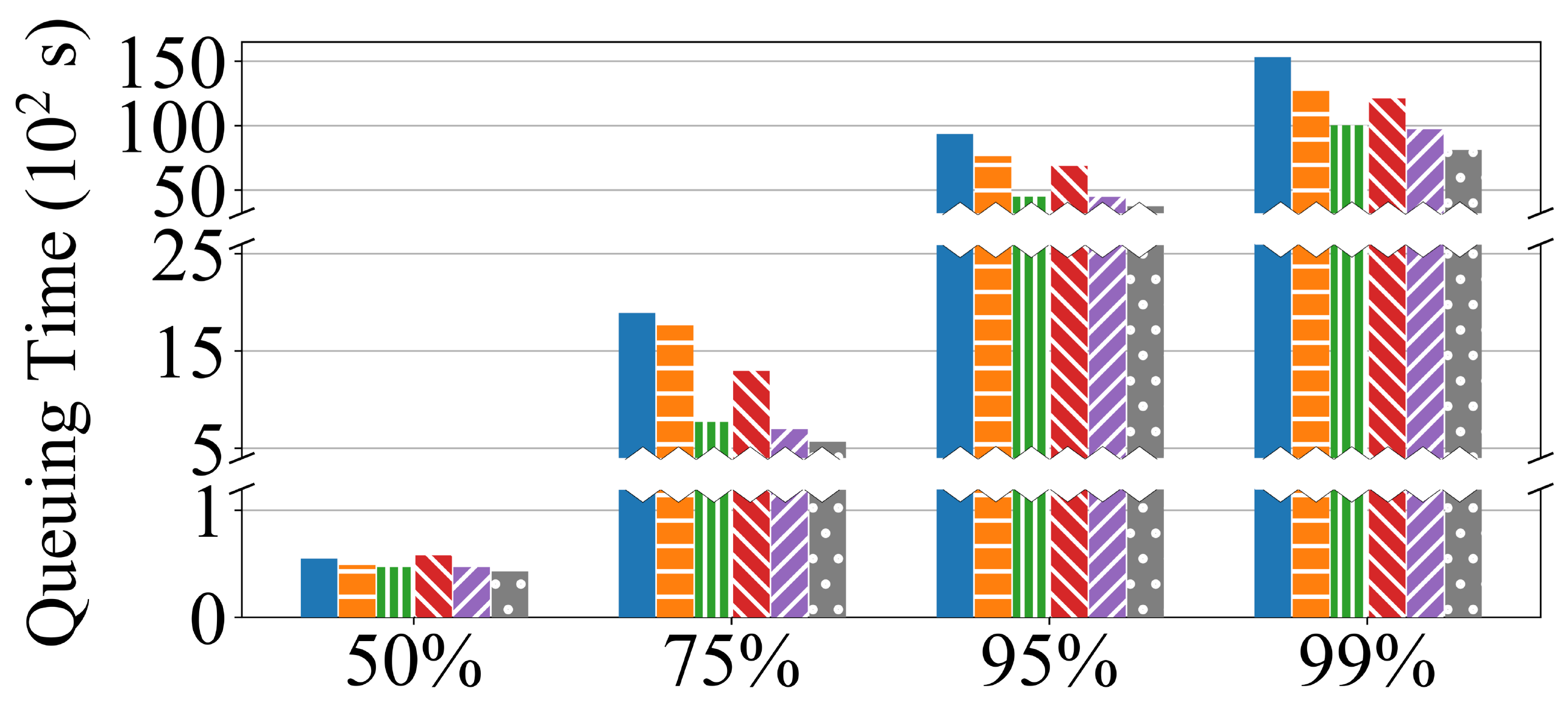}
	    \vspace{-7mm}
	    \captionof{figure}{50\%ile, 75\%ile, 95\%ile and 99\%ile of queuing time (Basic).}
	    \label{fig:evaluation_elastic_queuing_tile}
	\end{minipage}
	\hfill
\begin{minipage}[t]{0.245\textwidth}
    \centering
    \includegraphics[width=\linewidth]{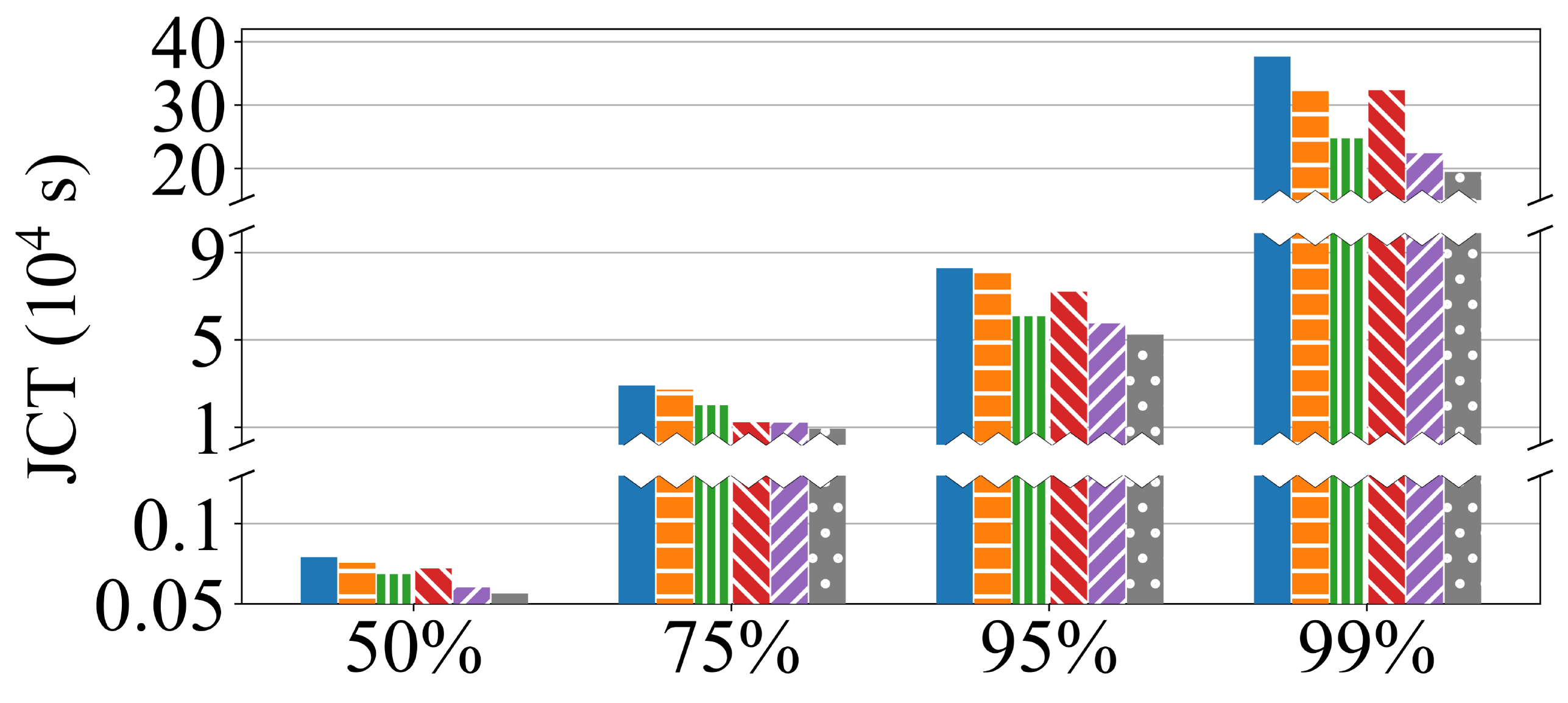}
    \vspace{-7mm}
    \captionof{figure}{50\%ile, 75\%ile, 95\%ile and 99\%ile of JCT (Basic).}
    \label{fig:evaluation_elastic_jct_tile}
\end{minipage}
\hfill
\begin{minipage}[t]{0.245\textwidth}
    \centering
    \includegraphics[width=1\linewidth]{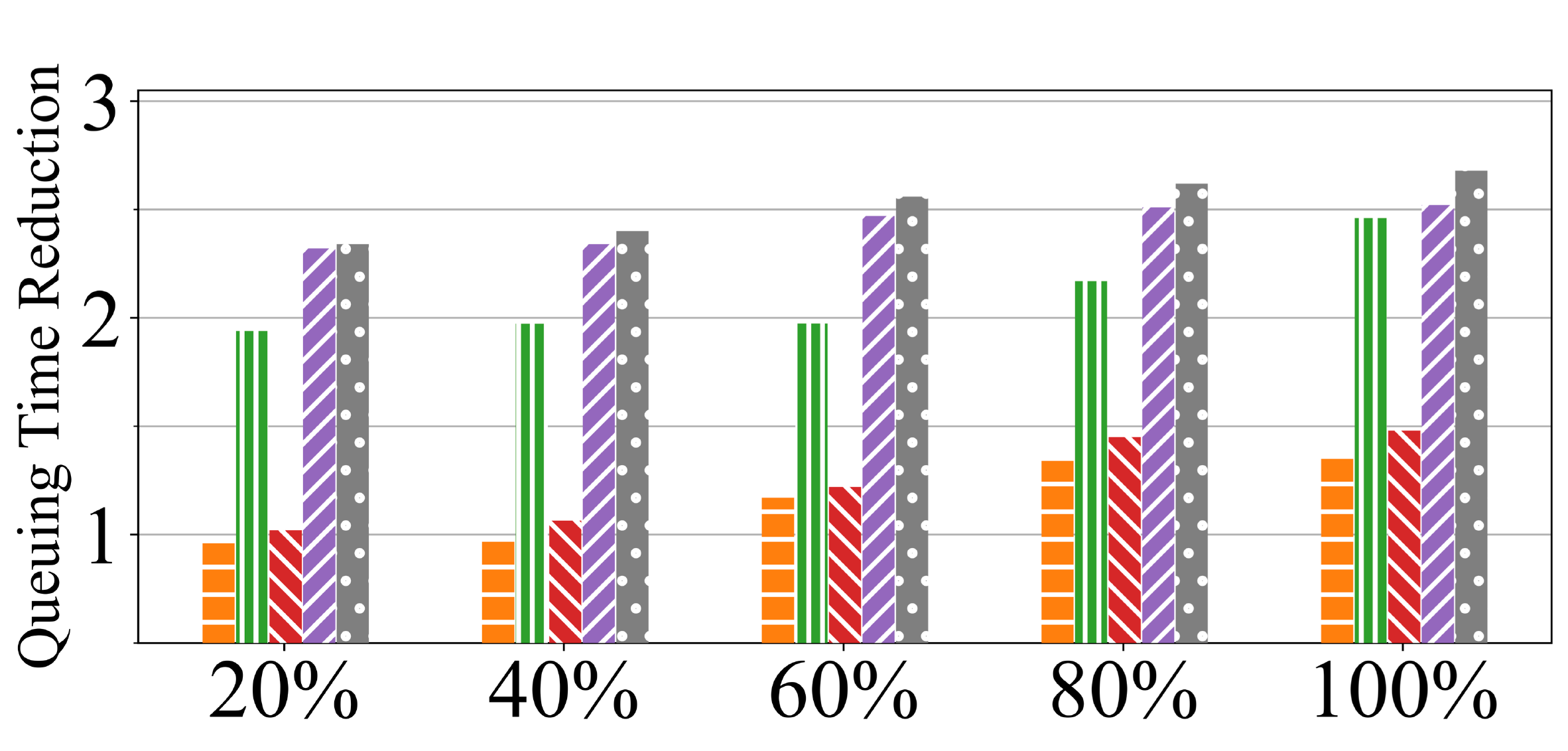}
    \vspace{-7mm}
    \caption{Queuing time reduction of Baseline as elastic jobs increases.}
\label{fig:percent_queuing}
\end{minipage}
\hfill
\begin{minipage}[t]{0.245\textwidth}
    \centering
    \includegraphics[width=1\linewidth]{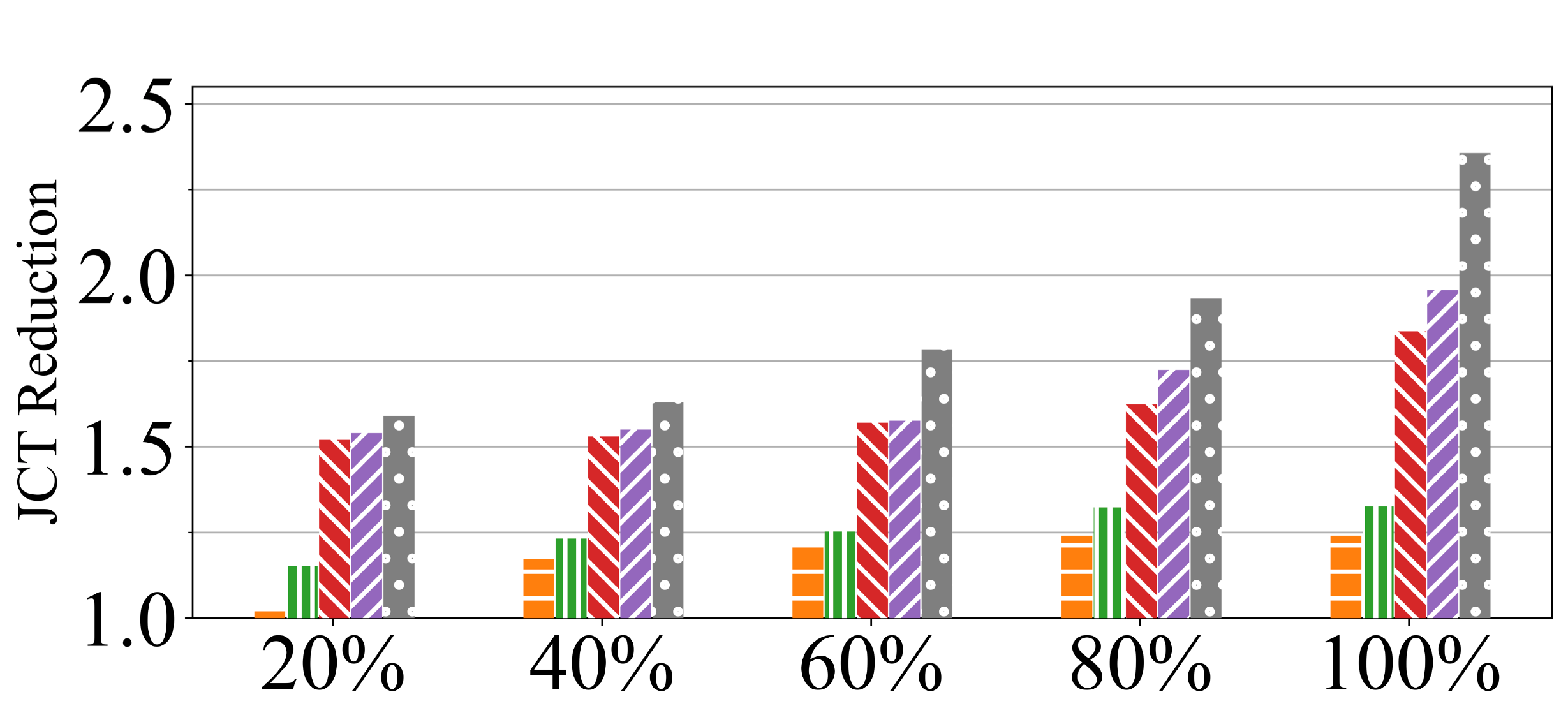}
    \vspace{-7mm}
    \caption{JCT reduction of Baseline as elastic jobs increases.}
\label{fig:percent_jct}
\end{minipage}
\vspace{-4mm}
\end{figure*}

\subsection{Deep-Dive: Capacity Loaning }
\label{sec:loaning_perf}
% We now dive into the two components of \sys.
We first focus on capacity loaning, aiming to understand its sources of gain and how our knapsack-based reclaiming heuristic (Algorithm~\ref{algo:reclaim}) compares to other schemes.
The results here are obtained without elastic scaling.

\noindent\textbf{Sources of gain.}
The JCT improvement mainly comes from reduction in queuing time as jobs now can run on the loaned resources instead of waiting in the queue.
Table~\ref{table:loan_jct} shows the statistics of queuing time and JCT for jobs running on the on-loan servers.
The median and 95\%ile queuing time is improved by 4.72x and 3.22x, respectively, compared to Baseline.
The resource usage rate of on-loan servers throughout the experiment is consistently above 93\% as depicted in Figure~\ref{fig:reclaim_usage}, which proves the effectiveness of resource loaning.
\noindent\textbf{Reclaiming heuristic.}
We compare our reclaiming heuristic to Random and SCF.
We consider two metrics,
percentage of preempted jobs among running jobs, and collateral damage as the fraction of GPUs vacated in excess of the reclaiming demand.
It is clear from Figure~\ref{fig:simulator_preemption} \sys outperforms
other solutions with and without elastic scaling.
Without scaling, \sys's knapsack-based heuristic reduces preemption and
collateral damage by 1.5x, 1.67x and 1.37x, 1.59x over SCF and Random, respectively.
With scaling, \sys scales elastic jobs on the flexible server group first which
further widens the gap.
From Table~\ref{table:simulator_stat}, it is clear that reducing preemptions
is beneficial: \sys reduces the average queuing time and JCT by 1.26x, 1.28x and 1.20x and 1.18x over SCF and Random.

We also run an exhaustive search to find the optimal reclaiming
solution as the upper bound.
\sys results in the same number of preemptions as optimal when reclaiming fewer than 60 servers, and incurs 19\% more preemptions otherwise. \sys's reclaiming decision shows an average 84\% resemblance as the optimal solution.
The average running time of the optimal solution, however, is 420k times that of
\sys.

\subsection{Deep-Dive: Job Scheduling }
\label{sec:scaling_perf}
We evaluate job scheduling in more details here.
The results are obtained without capacity loaning in Basic scenario.

\noindent\textbf{Sources of gain.}
Figures~\ref{fig:evaluation_elastic_queuing_tile}--\ref{fig:evaluation_elastic_jct_tile}
plot the distribution of queuing time and JCT for all schemes.
Our key insight in solving the scheduling problem is to prioritize
the inelastic workload (\cref{sec:greedy_allocation}).
Gandiva does not improve Baseline much due to its opportunistic nature: it only scales jobs in low-utilization periods.
Both \sys and AFS allocate the minimum demand to each job initially.
From Figure~\ref{fig:evaluation_elastic_queuing_tile}, they have similar median queuing time.
Though Pollux considers job's minimum demand and favors those with large goodput, it does not explicitly launch as many jobs as possible, thus incurring longer queuing time.  
\sys outperforms Pollux by 1.23x and 1.69x in median and 95\%ile queuing time.

Turning to JCT, we find from Figure~\ref{fig:evaluation_elastic_jct_tile} that Pollux tends to prolong the large-and-long jobs by shrinking their resources towards the end of training to yield for newly-started jobs that make rapid progress with the same resources.
% Jobs in Pollux exhibit a long JCT for 95\% to 99\% jobs in .
Moreover, Pollux's performance heavily hinges upon the problem scale and the number of iterations allowed for its genetic algorithm.
In a large cluster of over 3,500 GPUs with heavy workload, the preset 100 iterations
are not sufficient to get an efficient allocation result.
To keep the scheduling overhead acceptable, we set the number of iteration
to 250 and \sys still has 1.20x and 1.25x improvements in median and 95\%ile
JCT.
AFS assumes unbounded elasticity and shows a higher resource usage.
However, unlimited elasticity and greedy allocation implicitly favor jobs with better throughput at the cost of others.
Its average JCT is 1.2x that of \sys which balances the resources each job gets by making global allocation and considering limited elasticity.

\noindent\textbf{Sensitivity analysis: Proportion of elastic jobs.}
We wish to
analyze whether \sys is sensitive to the proportion of elastic jobs in the mix.
Figure~\ref{fig:percent_queuing} shows the performance comparison when elastic
jobs grow from 20\% to 100\% of the population.
All schemes show improvements as a result.
\sys delivers the largest gains in both queuing time and JCT compared to other schemes with more elastic jobs, demonstrating that its scheduler most efficiently exploits job elasticity.
AFS also has good gains in queuing time as it initially allocates minimum demand to each job.
Its JCT gains, however, are much lower due to the greedy heuristic in ordering the jobs for allocation.
Pollux's queuing time performance is poor as queuing time is not considered in its design. Its JCTs are much better because it auto-tunes the hyperparameters for best performance.

% \ljm
{
\noindent\textbf{Using hyperparameter tuning.}
% It is possible to tune the training hyperparameters such as batch size and learning rate dynamically when the resource allocation changes. Pollux shows that this is beneficial to job scheduling.
We study \sysP now which adapts Pollux's job agent to tune jobs' hyperparameters as explained in \Cref{sec:setup}. 
% in Figures~\ref{fig:evaluation_elastic_queuing_tile}--\ref{fig:percent_jct}.
% In this experiment, we use the DNN models given by the distribution in Pollux's paper \cite{qiao2020pollux} and adapt its job agent to \sys, denoted as \sysP in Figures~\ref{fig:evaluation_elastic_queuing_tile}--\ref{fig:percent_jct}. 
% In \sysP, Pollux's job agent searches the best hyperparameter and reports the estimated goodput to the job scheduler. 
In the Basic scenario, \sysP (row 12 in Table~\ref{table:simulator_stat}) contributes an additional 18\% and 13\% improvements over Baseline in 95\%ile JCT and 99\%ile JCT.
%  as job running time is further reduced with tuned hyperparameters.
The improvement is more significant when all the jobs
are elastic as seen in Figures~\ref{fig:percent_queuing}--\ref{fig:percent_jct}. 
% The average JCT is reduced by 1.21x over \sys, and it also leads to an additional 14\% gain in queuing time than \sys (Figures~\ref{fig:percent_queuing}--\ref{fig:percent_jct}).

More importantly, \sysP allows for a fair comparison of job scheduling against Pollux as both have hyperparameter tuning now.
It outperforms Pollux by 1.32x and 1.37x in median and 95\%ile JCT in Basic scenario (Figure~\ref{fig:evaluation_elastic_jct_tile}). 
% \sysP improves the average JCT by 1.29x and 1.18x over Pollux in 80\% and 100\% elastic jobs, respectively (Figure~\ref{fig:percent_jct}). 
% \ljm{
% In terms of average JCT, \sysP's gain over Pollux is 21\% larger than \sys's when elastic scaling is enabled for all the jobs,
% which shows our scheduling policy performs much better.
\sys's gain over Pollux is larger here which shows that \sys's scheduling policy performs better in JCT. 
% }
The main reason is that \sys specifically optimizes JCT while Pollux optimizes goodput in order to improve resource efficiency. Thus JCT for some jobs is affected especially near the end of training when the marginal gain of resources becomes smaller (i.e. goodput is lower) and resource allocation is decreased. 
Another side-effect of goodput-based scheduling is back-and-forth scaling as goodput varies as soon as hyperparameter or allocation changes. We find the total scaling times of Pollux is 1.76x that of \sysP in the Ideal scenario, and many are scaling-out followed immediately by scaling-in in the next interval. This may also degrade JCT. 
% . During the periods of cluster contention, 22.1\% of 
% scaling out operations are followed by scaling in within the 
% next scheduling interval. 

\begin{comment}
  
\ljm{
Pollux's allocation decisions are sensitive to change in job goodput. Its scheduler adjusts job resource allocation whenever the hyperparameters changes and vice versa.
Pollux need to search every combination of job resources and hyperparameter during scheduling as both 
scheduler and job agent are optimizing goodput.
We find that it frequently scales running jobs back and forth 
to improve cluster-wise goodput, which does not necessarily reduce the job running time.
In \sysP, job scheduler optimizes average JCT exclusively to 
achieve higher cluster efficiency and job agent responds 
the resource change passively to handle job-level goodput.
\sysP can reduce the total scaling operations and 
the back-and-forth scaling by 43\% and 36\%, respectively,  
as it solely considers job's JCT reduction value 
and always favors jobs with a larger JCT improvement. 
}
% \hxq{what's the relevance...}
% \hxq{we may need to revise the wording here, seems a bit too dismissive to Pollux}
}
\end{comment}

\subsection{Testbed Results}
\label{sec:testbed_results}

Here we use our prototype in testbed experiments to
schedule jobs and YARN to run, scale, and preempt them on servers.

\noindent\textbf{Workload.}
We use a scaled-down version of the
traces with 180 training jobs with 10 elastic ones {(similar to the Basic scenario)}; jobs with (maximum) demand
larger than 16 GPUs (50\% cluster) are excluded.
Job submission in the trace lasts for 8 hours and training time
varies from 2 minutes to 2 hours.
The inference trace is also scaled down according to the testbed capacity.

\noindent\textbf{JCT and queuing time.}
Table~\ref{table:testbed_queuing_jct} shows the statistics of queuing time and JCT.
\sys improves average and 95\%ile queuing time by 1.38x and 1.36x over Baseline (row group 1).
In terms of JCT, \sys improves the median and 95\%ile by 19.9\% and 11.7\% over
Baseline.
The gains come from both capacity loaning and elastic scaling: the orchestrator
performed 6 loaning and 8 reclaiming operations involving a total of 10
servers, and the scheduler issued 73 scaling operations.
In capacity loaning, \sys outperforms Random and SCF by 19\% and 15\% in average 
queuing time. In elastic scaling, \sys's tail queuing time is 10\% shorter than AFS. 
Its JCT gain is 1.19x over Baseline compared to 1.14x and 1.15x for AFS and Pollux. 

The results here show that \sys is highly effective in reducing queuing time.
The JCT improvements are relatively small due to the inference cluster's
limited resources compared to job demand.
We observe the inference cluster loaned at most three servers which is equivalent to one training server in computational capability, while it is common for a job to demand an entire training server in our trace.

\begin{table}[t]
  \centering
\resizebox{\columnwidth}{!}{
  \begin{tabular}{@{}lllllllll@{}}
    \toprule \multirow{2}{*}{Scenario}
    & \multirow{2}{*}{Solution} & \multicolumn{3}{c}{Queuing Time (s)} & \multicolumn{3}{c}{JCT (s)} & Preemption \\ \cmidrule(l){3-9} 
    &         & Mean     & Median     & 95\%ile     & Mean  & Median  & 95\%ile  &  \multicolumn{1}{c}{Ratio} \\ \midrule
    \multirow{2}{0.9cm}{Basic} 
    & Baseline & 1532 & 772 & 1003    & 4078 & 2183 & 3096 & 0\\
    & \sys & 1109 & 503 & 738 & 3335 & 1747 & 2731 & 18\%\\\midrule
    \multirow{3}{0.9cm}{Capacity Loaning} 
    & Random  &  1527    &  658      &  993        &  3893 &  2046   &  3015  & 34\%\\                         
    & SCF     &  1473    &  614      &  864        &  3857 &  1994   &  3001  & 30\% \\
    & \sys    &  1230    &  594      &  823        &  3748 &  1946   &  2864  & 22\% \\ \midrule
    \multirow{4}{0.9cm}{Elastic Scaling}  
    & Gandiva &  1443    &  645       &  1002      &  3882 &  2015   &  2893   & NA \\
    & AFS     &  1338    &  534       &  882       &  3521 &  1836   &  2803   & NA\\
    & Pollux  &  1405    &  576       &  937       &  3552 &  1934   &  3004   & NA\\
    & \sys    &  1318    &  546       &  798       &  3413 &  1791   &  2794  & NA \\ \bottomrule
    \end{tabular}}
  \vspace{-3mm}
  \caption{Testbed results using different solutions.}
  \label{table:testbed_queuing_jct}
  \vspace{-2mm}
  \end{table}

\noindent\textbf{Preemption.}
% We also investigate the job preemption in testbed.
% Figure~\ref{fig:testbed_preemption} shows the total number of preemptions and the corresponding collateral damage in testbed.
\sys reduces preemption significantly by over 1.3x compared to Random and SCF reclaiming schemes (row group 2).
We also measure the preemption overhead, including the time
to save a checkpoint to the disk, terminate containers, launch new containers on different servers,
and load the checkpoint before training starts.
The average overhead is 63 seconds, which is adopted in our large-scale simulation.

  \begin{figure}[t]
    \centering
    \includegraphics[width=0.65\linewidth]
    {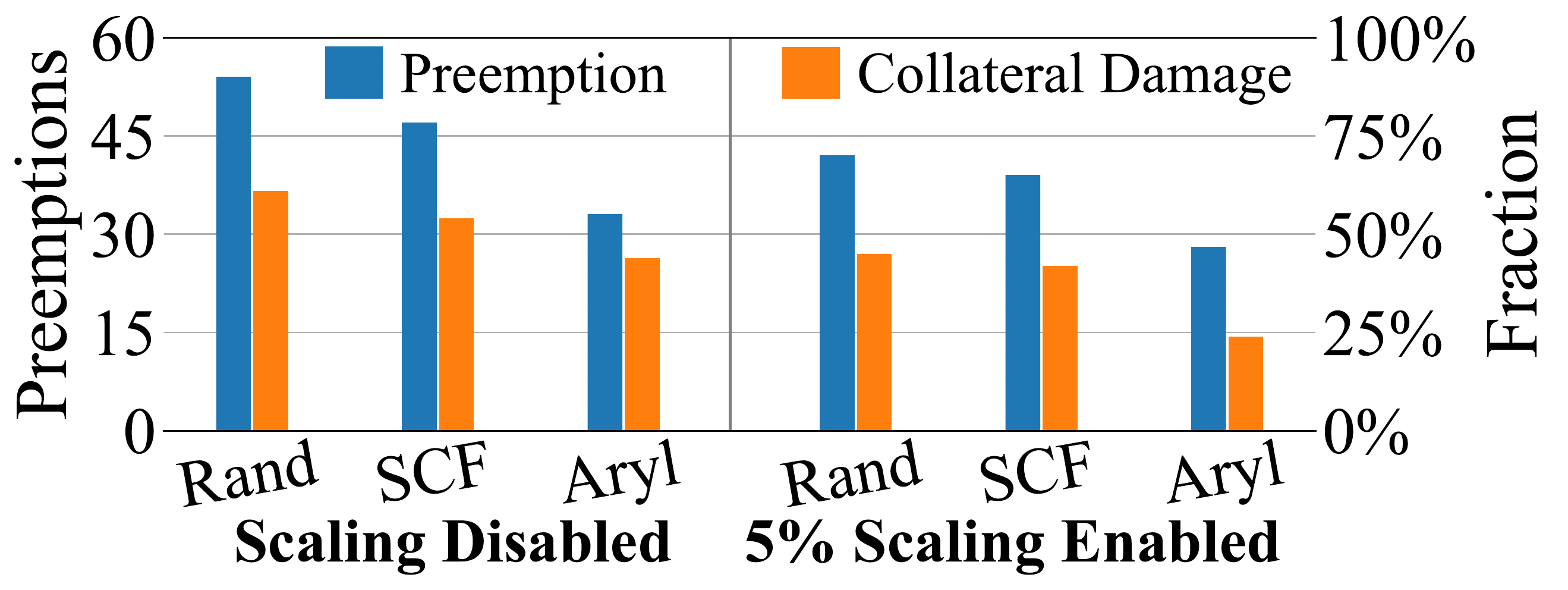}
    \vspace{-3mm}
    \caption{The number of job preemptions and average collateral damage comparison in testbed. }
    \label{fig:testbed_preemption}
    \vspace{-4mm}
\end{figure}
\noindent\textbf{Sensitivity analysis: Error in running time estimation.}
% \ljm{TODO: move to appendix in ATC version}
Our second sensitivity analysis concerns the running time prediction which
\sys's scheduler relies on.
Table~\ref{table:inaccurate_profiler} shows the performance under
different estimation accuracy.
\sys improves queuing delay by 1.76x over Baseline even
when there are 60\% wrong predictions (each with at most 25\% error).
Its gain is consistent with less than 60\% wrong predictions.
% This demonstrates that \sys is robust to prediction error in running time.

\begin{table}[t]
\centering
      \resizebox{0.6\columnwidth}{!}{
      \begin{tabular}{@{}ccc@{}}
      \toprule
      \% wrong prediction & Queuing time gain & JCT gain \\ \midrule
      0\%              & 2.37                    & 1.57            \\
      20\%             & 2.21                     & 1.52            \\
      40\%             & 2.17                     & 1.49            \\
      60\%             & 1.76                      & 1.38            \\ \bottomrule
      \end{tabular}}
      \vspace{-2mm}
      \caption{Queuing delay and JCT gain with incorrect running time estimation.
      The fraction of incorrect estimation varies from 0\% to 60\%.
      We assume each incorrect prediction is within an error margin of 25\%.
      }
      \label{table:inaccurate_profiler}
    % \end{minipage}
    \end{table}

%% file: discussion.tex
%!TEX root = main.tex
% \vspace{-3mm}
\section{Discussion}
\label{sec:discussion}
% \vspace{-2mm}
% We discuss some issues pertaining to \sys's design.
% 

\noindent\textbf{Fine-grained resource sharing.}
\sys uses bare-metal machine as the basic unit of loaning and reclaiming.
Our intention is to avoid interference between training and inference.
This concern can be alleviated by improvements from the infrastructure
(e.g. better isolation mechanisms).
Then one may consider fine-grained sharing on the GPU level, which allows more sharing opportunities but also demands a
more careful scheduling design because of the larger problem scale.

\noindent\textbf{Performance under scaling.}
We assume the elastic job's training throughput is linear in the amount
of allocated resources within the scaling range.
In practice training throughput is likely to scale sub-linearly due to factors such as network communication and synchronization overhead.
An improved approach may be to empirically profile the throughput and running time of the workloads as a non-linear function of resources.
\sys's scheduling algorithm still works with non-linear scaling which does not
change the combinatorial nature of the problem; we provided simulation results
in~\cref{subsec:simulation}.

% \noindent\textbf{Flexible worker placement.}
% Our placement solution packs flexible workers together on a subset of
% servers, and is effective in minimizing preemptions.
% Yet, it neglects the communication cost when workers of the same job are
% located on different servers.
% Though network bandwidth is under-utilized in most cases, further optimization
% that minimizes the network distance of the workers is still beneficial in
% improving the training efficiency.

\noindent\textbf{Heterogeneous GPU training.}
Training with heterogeneous GPUs is still an active area of research and the current mechanisms are primitive \cite{chen2020semi}.
We observe that though adjusting the batch size can roughly synchronize the
workers, it may prolong the convergence of the model in some cases.
More effort is needed to improve training efficiency with heterogeneous GPUs and to automate hyperparameter adjustment \cite{chaudhary2020balancing,258896}.

% \noindent\textbf{Elastic Scaling Jobs.} The scaling requirement of DL elastic jobs
% are still rigid, which consists of scale range constraints, worker-unit scaling.
% The existing design though provides stability for training jobs, is still a
% challenge for job schedulers. Scheduling policy is complicated by these factors
% and resource fragmentation persist. Further investigations on how to control
% the job scaling with more flexibility are required.

\begin{comment}
\noindent\textbf{Running time estimation.} Although \sys improves the average
JCT and cluster resource usage even with some wrong predictions of running time,
running time estimation is necessary to make resource allocation decisions on
elastic scaling jobs. It is difficult to evaluate how much a job can improve with
more resources without running time information. The job profiler we build is naive
and could lack generality.
Job running time profiler with stable prediction accuracy will benefit \sys.
\end{comment}

%% file: related.tex
%!TEX root = main.tex
\section{Related Work}
\label{sec:related}
We now discuss related work not mentioned in \cref{sec:motivation}.

\noindent\textbf{GPU cluster schedulers.}
There are some schedulers tailored for GPU training clusters.
We have discussed Pollux, AFS, and Gandiva extensively in \cref{sec:comparison} and \cref{subsec:simulation}.
Tiresias~\cite{gu2019tiresias} applies least-attained-service to minimize average JCT.
%  and relax the locality constraints without hurting the job performance.
It does not consider elastic scaling.
Optimus\cite{peng2018optimus} schedules jobs with a
online fitting model, which predicts training model's running time.
% Predicting a DNN's convergence however is challenging as discussed in \cite{he2016deep}. 
PAI~\cite{weng2022mlaas} introduces a 
scheduler which reserves high-end GPUs for high-GPU tasks and packs low-GPU tasks on less advanced GPUs. 
% Gandiva\cite{xiao2018gandiva} optimizes resource utilization
% through fine-grained time and space sharing.
% Pollux\cite{qiao2020pollux} introduces a scheduler jointly optimizing cluster-wide resource allocation and hyperparameters of DNN jobs simultaneously.
% AFS\cite{265013} proposes a scheduling algorithm greedily allocating more resources to the most efficient jobs.
% Though these works consider the existence of elastic jobs, they neglect
% the practical constraints of scaling DNN jobs.
These works all schedule jobs in a cluster with fixed capacity.
In~\Cref{app:scheduler_detail_comparison}, we further elaborate the differences
between \sys and existing DNN schedulers.

\noindent\textbf{Systems support for elastic scaling.}
There is emerging interest in exploiting resource elasticity in distributed
training.
Systems such as~\cite{elasticdl:sql,
ElasticHorovod:Website, PyTorchElasticLaunch:Website} extend various ML
frameworks to support elasticity.
\cite{or2020resource} proposes an auto-scaling policy by considering 
cost and scaling efficiency and proves that the scaling overhead is only 4\% of checkpointing overhead. 
% It identifies that scaling efficiency does not improve linearly with more
% resource allocated.
% It introduces a scaling condition based on its marginal utility and cost.
AntMan~\cite{258957} provides a scaling mechanism to micromanage computation and GPU memory during training, and a job scheduler for performance guarantees.
They are complementary to \sys as they provide practical solutions for
scaling DNN jobs. % as they provide practical scaling systems. I do not understand the meaning so i comment it out [guo]

\noindent\textbf{Dynamic resource allocation.}
Graphene \cite{grandl2016graphene} and PriorityMeister \cite{zhu2014prioritymeister} dynamically adjust resource allocation to
fit job's time-varying demand and utilize resources more efficiently.
In \sys, we consider scaling for jobs that can work with a range of resources, which are taken as %inequality %I do not understand what the meaning so i comment it out.
constraints to the scheduling
problem.
\sys schedules jobs with an extra dimension of how much resource should a job get
and its impact on cluster performance.

\begin{comment}
\noindent\textbf{Heterogeneous GPU scheduling.}
Allox \cite{le2020allox}, Gandiva\textsubscript{fair}
\cite{chaudhary2020balancing}, and Gavel \cite{258896}
are purposely built to support heterogeneous GPU training.
Allox aims at minimizing average JCT in a CPU-GPU hybrid cluster. It considers the resource interchangeability of jobs to decide
whether a job runs on CPU or GPU.
Gandiva\textsubscript{fair} introduces a heterogeneity-agnostic max-min fairness policy to improve cluster efficiency.
Gavel proposes a round-based preemptive scheduling algorithm where in each round jobs might be preempted and re-launched at a different GPU to improve the cluster-wise throughput.
These solutions do not consider capacity loaning or elastic scaling.
\end{comment}

%% file: appendix.tex
%!TEX root = main.tex
% \documentclass[letterpaper,twocolumn,10pt]{article}
% \usepackage{usenix2019_v3}
% \usepackage{enumitem, kantlipsum}
% \usepackage{amsmath,comment,adjustbox}
% \setlist[enumerate]{itemsep=0mm}
% \begin{document}

% \title{Appendices}
% \maketitle

\clearpage
\appendix
\section{Heterogeneous GPU Training}
\label{app:heterogeneous}
\begin{table*}[t]
  \centering
  \resizebox{14cm}{!}{
  \begin{tabular}{|p{4.5cm}|p{1cm}|p{1cm}|p{1cm}|p{1cm}|l|l|l|}
  \hline
  \multirow{2}{*}{Models \& Dataset \& Metric} & \multicolumn{2}{c|}{Allocation} & \multicolumn{2}{c|}{Batch Size} & \multicolumn{2}{c|}{Throughput} & \multicolumn{1}{c|}{\multirow{2}{*}{Running Time (m)}} \\ \cline{2-7}
                                               & V100            & T4            & V100            & T4            & Theoretical    & Experimental   & \multicolumn{1}{c|}{}                              \\ \hline\hline
  \multirow{4}{*}{\shortstack[l]{ResNet-50\cite{he2016deep} \& ImageNet\cite{deng2009imagenet} \\ \& Top 1 Accuracy: 75\%}}
                                               & 16              & 0             & 256             & -             & -              & 18883          &  127                                             \\ \cline{2-8}
                                               & 12              & 4             & 248             & 85            & 15342          & 14383          &  162                                               \\ \cline{2-8}
                                               & 8               & 8             & 250             & 84            & 11801          & 9823           &  155                                               \\ \cline{2-8}
                                               & 4               & 12            & 237             & 82            & 8261           & 7297           &  161                                                 \\ \hline\hline
  \multirow{4}{*}{\shortstack[l]{GNMT\cite{wu2016google} \& WMT'16\cite{wmt} \\ \& BLEU: 25.5}}
                                               & 16              & 0             & 128             & -             & -              & 791771         &  35                                                \\ \cline{2-8}
                                               & 12              & 4             & 128             & 66            & 643313         & 484684         &  46                                                \\ \cline{2-8}
                                               & 8               & 8             & 123             & 67            & 494856         & 436631         &  47                                                \\ \cline{2-8}
                                               & 4               & 12            & 124             & 66            & 346399         & 312033         &  44                                                \\ \hline\hline
  \multirow{4}{*}{\shortstack[l]{BERT\cite{devlin2018bert} \& SQuaD v1.1\cite{rajpurkar2016squad} \\ \& F1: 86}}
                                               & 32              & 0             & 10              & -             & -              & 1572           &  46                                                \\ \cline{2-8}
                                               & 24              & 8             & 10              & 4             & 1277           & 923            &  61                                                \\ \cline{2-8}
                                               & 16              & 16            & 9               & 5             & 982            & 874            &  61                                                \\ \cline{2-8}
                                               & 8               & 24            & 9               & 5             & 687            & 655            &  57                                                \\ \hline

  \end{tabular}}
  \centering
  \begin{tablenotes}
   \scriptsize
   \item (1) Theoretical throughput is computed based on the computation capacity
   difference with homogeneous training using V100 GPU. Here, we consider T4 to be one fourth of V100.
  \end{tablenotes}
  \vspace{-2mm}
  \caption{Performance of DNN jobs adopting heterogeneous training. Local batch size
  is measured when it is stabled. }
  \vspace{-2mm}
  \label{table:heterogeneous_training}
\end{table*}
In \cref{sec:loaning}, we consider heterogeneous training as a special approach
to utilize different types of GPUs. We also evaluate \sys in a scenario
where 10\% of jobs can run on heterogeneous GPU with non-ideal performance in~\cref{subsec:simulation}.
We now provide some details regarding the implementation of heterogeneous GPU
training and empirical analysis of the training performance.
Distributed data-parallel training with bulk synchronous parallel for communication
is adopted as they are the most widely used techniques in distributed training.
The key to efficient training is to balance the training time among workers and avoid stragglers.
We adopt a similar approach as \cite{chen2020semi} to tune the batch size of each
work in an online manner.
Initially, we set the local batch size of each worker based on their
computation capacity and GPU memory constraints. Specifically, the batch size
of T4 workers is one-fourth of the batch size of V100 workers.
During training, we adjust the local batch size and ensure a similar training
time for each iteration.
We initialize the window size to be 10 and the step size is set to 1.
At each observation window, local batch size of leader workers will
increase and that of straggler workers will decrease in step size.

Table~\ref{table:heterogeneous_training} shows the throughput
and running time of different DNN jobs running on V100 and T4.
By comparing the experimental results with the theoretical throughput, we find
naively tuning batch size without additional optimization techniques leads
to an average degradation of 27\%. To achieve the same validation metric,
the running time is also lengthened by up to 34\% compared with homogeneous training
with V100. Therefore, we quantify the non-ideal performance of heterogeneous training
to be 70\% of the theoretical results in~\cref{subsec:simulation}.

\section{Reclaiming Heuristics}
\label{app:reclaim_select}
Algorithm~\ref{algo:reclaim} shows the pseudo code of how to select servers
during reclaiming operation.
\begin{algorithm}[t]
  \caption{Server selection for reclaiming. }
  \label{algo:reclaim}
  \begin{algorithmic}[1]
    \Require $\mathcal{J}$: set of jobs on on-loan severs, where each job
    $j$ has info of $\mathcal{S}_j$, the set of servers it is hosted;
    $OnLoanList$: set of on-loan severs, where each server $s$ has info of
    $\mathcal{J}_s$, the set of jobs running on it;
    $N_R$: reclaim demand.
    % \Statex $OnLoanList$: Set of on-loan severs, $D$: reclaim demand.
    \Ensure $ReclaimList$, set of servers to reclaim
    \Procedure{InitPreemptionCost}{$\mathcal{J}, OnLoanList$}
        \State $Q \gets$ Queue() %\Comment{server and its cost}
        \For{server $s \in OnLoanList$}
                \State $Q$.push$(\langle s, 0\rangle)$ \Comment{initialize preemption cost}
        \EndFor
        \For{job $j \in \mathcal{J}$}
                    \For{$s \in \mathcal{S}_{j}$}
                        \State $Q$.add\_preemption\_cost$(s, {1}/
                        {\left|\mathcal{S}_j\right|})$ \label{reclaim:add}
                        \Comment{add job cost to $s$}
                    \EndFor
            \EndFor
        \State SortByCostIncreasing$(Q)$
        \State \textbf{return} $Q$
    \EndProcedure
    \Procedure{UpdateCost}{$Q, s'$}
        \For{job $j \in$ $\mathcal{J}_{s'}$}
            \For{$s \in \mathcal{S}_{j}$ and $s\notin s'$}
                    \State $\mathcal{J}_s$.remove\_job$(j)$ \Comment{preempt job
                    $j$}
                    \State $Q$.subtract\_preemption\_cost$(s, {1}/
                    {\left|\mathcal{S}_j\right|})$
                \EndFor
        \EndFor
        \State SortByCostIncreasing$(Q)$ \label{reclaim:sort}
        \State \textbf{return} $Q$
    \EndProcedure
    \Procedure{SelectServers}{$\mathcal{J}, OnLoanList, N_R$}
        \If{$N_R = 1$}
            \State $s \gets$ FindBestServer$(OnLoanList)$
            \State \textbf{return} ReclaimList$(s)$
        \EndIf
        \State Q $\gets$ InitPreemptionCost$(\mathcal{J}, OnLoanList)$
        \State $ReclaimList \gets$ []
        \While{$N_R > 0$}
            \State $\langle s, cost\rangle \gets Q$.top$()$
            \State $ReclaimList$.append$(s)$
            \State $Q$.pop$()$ \Comment{remove from the top}
            \State $N_R \gets N_R - 1$
            \State UpdateCost$(Q, s)$ \label{reclaim:update}
        \EndWhile
     %    \Statex \Comment{The following are used to reduce idle servers, could be deleted and described in text}
     %    \State $redundant \gets 0$
     %    \State $RelatedServer \gets set()$
      % \For{server $s \in Q$}\Comment{in case redundant servers exist}
      %    \If{$s == 0$ and $s$ no $\in ReclaimList$}
      %        \State $redundant \gets redundant + 1$ \Comment{count redundant idle servers}
      %        \State $s_{relate} \gets$ GetRelateServer$(s)$
      %        \State $RelatedServer$.add$(s_{relate})$
      %    \EndIf
      % \EndFor
      % \For{server $s \in ReclaimList$}
      %   \If{$redundant > 0$ and $s$ not $\in RelatedServer$} \Comment{if $s$  has no relate server}
      %     \State $ReclaimList$.remove$(s)$
      %     \State $redundant \gets redundant - 1$
      %   \EndIf
      % \EndFor
      \State \textbf{return} $ReclaimList$
    \EndProcedure

  \end{algorithmic}
\end{algorithm}

\section{Analysis of Scheduling Elastic Jobs}
\label{app:elastic_theoretical_analysis}
In \cref{sec:scheduling_challenge}, we design a simplified example to show the
challenges of finding optimal resource allocation to elastic jobs. Here we provide
a formal analysis on which factors affect the optimal allocation in a
two-job scenario.

\noindent\textbf{Setup. } In a cluster with $C$ available GPUs,
two pending elastic jobs $p$ and $q$ are waiting for resource allocation.
Both jobs have specified their minimum and maximum resource demand.
Table~\ref{table:notation_analysis} lists the notations to be used in the
following analysis.

\begin{table}[h]
  \centering
  \resizebox{0.85\linewidth}{!}{\begin{tabular}{|l||p{7cm}|}
    \hline
    \textbf{Term}    & \textbf{Description}                             \\ \hline\hline
    $C$              & Number of GPUs in the cluster                    \\ \hline
    $L_j$            & Workload of job $j$                              \\ \hline
    $g_{min}(j)$     & Minimum (base) demand GPU of job $j$             \\ \hline
    $g_{max}(j)$     & Maximum demand GPU of job $j$                    \\ \hline
    $g_{j}$          & Allocation of job $j$                            \\ \hline
    $rt(L_{j}, g_{j})$      & Running time of job $j$ in $g_{j}$ allocation when
    the remaining workload is $L_{j}$    \\ \hline
  \end{tabular}}
  \vspace{-2mm}
  \caption{Notations and their descriptions. }
  \vspace{-6mm}
\label{table:notation_analysis}
\end{table}

\noindent\textbf{Objective. }The objective is to determine the resource allocation
$g_{p}$ and $g_{q}$
of the two elastic jobs $p$ and $q$ so that the average JCT is minimized.

\noindent\textbf{Definitions and constraints. }
To facilitate understanding, we still assume linear scalability of elastic jobs.
In specific, we define the workload $L$ to be the GPU hours of a job. For
elastic scaling jobs, the total workload is constant. Its running time can
be computed from $L$ and its resource allocation.
\begin{equation}
  rt(L, g_{min}) = \frac{L}{g_{min}}
\label{eqn:workload_def}
\end{equation}

We also make assumptions on the cluster capacity. We omit some of the scenarios
as they have straightforward results, including (1) the available resources
can merely fulfill the minimum demand of one job (2) there are abundant available
resources to host the maximum demand of both jobs. The most intricate case
is considered by adding the following constraints:
\begin{equation}
  \label{eqn:assumption}
  \begin{aligned}
    &  g_{max}(p) \leq g_{max}(q) < C\\
    &  g_{min}(p) + g_{min}(q) < C < g_{max}(p) + g_{max}(q)
  \end{aligned}
\end{equation}
where $C$ denotes the cluster capacity and it is constrained by the minimum
demand and maximum demand of pending jobs, implying that the cluster has
sufficient capacity to host two jobs simultaneously, but not enough to
host at their maximum demand.

\noindent\textbf{Problem formulation. } Since in Equation~\ref{eqn:assumption} we
already narrow down the value of cluster capacity, it can be inferred that
neither of the jobs will experience any queuing time.
We can then formulate the average JCT by solely considering their running time.
From Equation~\ref{eqn:workload_def}, the job running time can be derived from
its resource allocation. The average JCT can be represented as:

\begin{subequations}
  \label{eqn:formulation}
  \begin{align}
    \min_{g_{p}, g_{q}} \quad & f(g_{p}, g_{q}) && \tag{\ref{eqn:formulation}}\\
    \textrm{s.t}        \quad & g_{p} + g_{q} = C \label{eqn:formulation:c1}\\
                              & g_{min}(p) \leq g_{p} \leq g_{max}(p)  \label{eqn:formulation:c2}\\
                              & g_{min}(q) \leq g_{q} \leq g_{max}(q) \label{eqn:formulation:c3}
  \end{align}
\end{subequations}
To minimize the average JCT, it is reasonable to let the jobs reserve as many
GPUs as possible (Equation~\ref{eqn:formulation:c1}). Meanwhile, the resource
allocation should conform with the scaling constraints of the jobs
(Equation~\ref{eqn:formulation:c2}, \ref{eqn:formulation:c3}).

Under the initial allocation, the average JCT could be represented as:
\begin{equation}
  f(g_{p}, g_{q}) = \frac{1}{2}\times \sum_{i \in \{p, q\}}{\frac{L_{i}}{g_{i}}}
  \label{eqn:initial_jct}
\end{equation}
It is certain that either (1) two jobs complete at the same time or (2)
one of the jobs completes before the other.
In terms of the latter case, the uncompleted elastic job can
use the vacated resources and scale up to its maximum demand to shorten its running
time.
For example, job $p$ completes first before job ${q}$. The JCT can be computed as:
  \begin{align*}
    & JCT_{p} = rt(L_{p}, g_{p})  \\
    & JCT_{q} = JCT_{p} + rt(L_{q} - JCT_{p} \times g_{q}, g_{max}(q))
  \end{align*}
where the running time of job $q$ consists of two parts: (1) the time trained in
$g_{q}$ GPUs, which is the same as the running time of job $p$ and (2) the
time to train the remaining workload in its maximum demand. We could also derive
the JCT if job $q$ completes first in the same approach.

We refine the formulation case by case with details of the prerequisites.

\noindent\textbf{Case \rom{1}:} Job $p$ completes first.
By substituting $g_q$ with $C - g_p$, the average JCT can be further transformed as:
\begin{subequations}
  \label{eqn:pformulation}
  \begin{align}
    \min_{g_{p}} \quad & \frac{1}{2} \times (\frac{L_p}{g_{p}} + \frac{L_p}{g_{p}} + \frac{L_q - \frac{L_p}{g_{p}} \times (C -g_{p})}{g_{max}(q)}) && \tag{\ref{eqn:pformulation}}\\
    \textrm{s.t}        \quad & \frac{L_p}{g_{p}} < \frac{L_q}{C - g_{p}}, \label{eqn:pformulation:c1}\\
                              & g_{min}(p) \leq g_{p} \leq g_{max}(p),  \label{eqn:pformulation:c2}\\
                              & C - g_{max}(q) \leq g_{p} \leq C - g_{min}(q) \label{eqn:pformulation:c3}
  \end{align}
\end{subequations}
where Equation~\ref{eqn:pformulation:c1} guarantees job $p$ completes first in
the initial allocation (or
at the same time as job $q$).
Equation~\ref{eqn:pformulation:c2} and~\ref{eqn:pformulation:c3} constrain
the resources allocated to each job.

\noindent\textbf{Case \rom{2}:} Job $q$ completes first. Similar as Case (\rom{1}),
the average JCT is:
\begin{subequations}
  \label{eqn:qformulation}
  \begin{align}
    \min_{g_{p}} \quad & \frac{1}{2} \times (\frac{L_q}{C - g_{p}} + \frac{L_q}{C - g_{p}} + \frac{L_p - \frac{L_q}{C - g_{p}} \times g_{p}}{g_{max}(p)}) && \tag{\ref{eqn:qformulation}}\\
    \textrm{s.t}        \quad & \frac{L_q}{C - g_{p}}  <  \frac{L_p}{g_{p}}, \label{eqn:qformulation:c1}\\
                              & g_{min}(p) \leq g_{p} \leq g_{max}(p),  \label{eqn:qformulation:c2}\\
                              & C - g_{max}(q) \leq g_{p} \leq C - g_{min}(q) \label{eqn:qformulation:c3}
  \end{align}
\end{subequations}
where the remaining workload of $p$ is completed by $g_{max}(p)$ GPUs.

Equation~\ref{eqn:pformulation} and~\ref{eqn:qformulation}
intersect when two jobs completes at the same time (i.e. $\frac{L_p}{g_p} = \frac{L_q}{g_q}$).
With further simplification, the average JCT can be represented as:
\begin{equation}
  \label{eqn:simplified_jct}
f(g_{p})=
\begin{cases}
               \frac{L_p+L_q}{2\times g_{max}(q)} + \frac{L_p}{g_p} \times (1 - \frac{C}{2\times g_{max}(q)}) & \text{(\rom{1})}\\
               \frac{L_p+L_q}{2\times g_{max}(p)} + \frac{L_q}{C - g_p} \times (1 - \frac{C}{2\times g_{max}(p)}) & \text{(\rom{2})}\\
\end{cases}
\end{equation}
By integrating the constraints of both cases with the problem setup
(Equation~\ref{eqn:assumption}), we find the cluster capacity $C$
is the deciding factor as it changes the sign of the coefficient of $g_p$ (i.e. $1 - \frac{C}{2\times g_{max}(q)}$).

When the cluster capacity is within the following range:
\begin{equation}
  2 \times g_{max}(p) \leq C <  g_{max}(p) + g_{max}(q)
\end{equation}
which makes the value of Equation~\ref{eqn:simplified_jct} monotonically
decrease with $g_{p}$. The minimum average JCT occurs when $g_{p}$ is
maximum (i.e. fulfulling job $p$'s maximum demand).

However, when the cluster capacity is smaller as:
\begin{equation}
  g_{min}(p) + g_{min}(q) \leq C <  2 \times g_{max}(p)
\end{equation}
the value of Equation~\ref{eqn:simplified_jct} increases first and then decreases,
implying the minimum average JCT are at the end points of $g_p$'s interval.
Specifically, the optimal allocation is when job $p$ receives its maximum demand
or job $q$ receives its maximum demand, depending on the job workload.

We summarize the optimal allocation as follows.
\begin{itemize}[leftmargin=*]
  \item When the cluster capacity $C \in [2 \times g_{max}(p), g_{max}(p) + g_{max}(q)]$,
fulfilling the maximum demand of job $p$ (i.e. the job with a smaller $g_{max}$) will bring best average JCT.
  \item When the cluster capacity $C \in [g_{min}(p) + g_{min}(q), 2 \times g_{max}(p)]$,
  the best average JCT is when the job with a smaller workload $L$ receives its maximum demand.

\end{itemize}
\noindent\textbf{Conclusion. }
From the result, we find that the optimal
resource allocation is affected by multiple factors, including cluster capacity
$C$, the job workload $L$ and the scaling range $g_{min}, g_{max}$.
Given the complexity of scheduling two elastic jobs, generalizing a consistent
solution of optimal resource allocation when there are additional constraints
(e.g. inelastic jobs, dynamic cluster, GPU demand per worker) is hard.
Therefore, we propose a two-phase heuristic that prioritizes base demand to
reduce queuing time and resorts to Knapsack packing to shorten the running time
of elastic jobs.

\section{Job Scheduling Heuristics}
\label{app:scaling_allocation}
Algorithm~\ref{algo:elastic} shows the detailed steps of how we schedule and
allocated resources to the queuing jobs.
\begin{algorithm}[t]
  \caption{Two-phase heuristic for resource allocation}
  \label{algo:elastic}
  \begin{algorithmic}[1]
    \Require $\mathcal{J}^q$: set of queuing jobs,
    $\mathcal{J}_e^r$: set of running elastic jobs, $\quad$
    ${C}^a$: available resources
    % \Statex $OnLoanList$: Set of on-loan severs, $D$: reclaim demand.
    %\Ensure Resource allocation and placement decision
    \Procedure{SortJobs}{$\mathcal{J}$}
        \State $\mathcal{J'} \gets$ []
        \For{job $j \in \mathcal{J}$}
            \If{$j$ is elastic}
                \State $\mathcal{J'}$.append$(\langle j, T^{max}_j\rangle)$ \Comment
                {max. running time}
            \Else
                \State $\mathcal{J'}$.append$(\langle j, T_{j}\rangle)$
            \EndIf
        \EndFor
        \State $\mathcal{J'} \gets$ SortByRunTimeIncrease$(\mathcal{J'})$
        \State \textbf{return} $\mathcal{J'}$
    \EndProcedure
    \Procedure{AllocateElastic}{$\mathcal{J}$, $C$}
        \State $groups \gets$ [] \Comment{multiple-choice knapsack}
        \For{job $j \in \mathcal{J} $}
            \State $g \gets$ NewGroup()
            \State $groups$.append$(g)$
              \For{$w=1$; $w\leq w^{max}_j-w^{min}_j$; $w++$}
                  \State $m \gets$ NewItem$()$
                  \State $m.weight \gets (w + w^{min}_j) D_j$ \Comment
                  {$D_j$: per-worker  GPU demand}
                  \State $m.value \gets T^{max}_j\times  w/(w+w^{min}_j)$
                  % \Comment{running time reduction}
                  \State $groups$.append$(m)$
              \EndFor
        \EndFor
        \State $\mathcal{S} \gets$ MaxValueDP$(groups, C)$ \Comment{allocation
        result}
        \State \textbf{return} $\mathcal{S}$
    \EndProcedure
    \Procedure{Allocation}{$\mathcal{J}^q$, $\mathcal{J}^r_e$, $
    \mathcal{C}^a$}
        \State $\mathcal{J} \gets$ SortJobs$(\mathcal{J}^q)$
        \State $\mathcal{J}^*, {C}^* \gets$ AllocateInelastic$(\mathcal{J},
        {C}^a)$ \Comment{SJF for inelastic demand}
        \If{${C}^* \geq 0$ }
            % \State $RJ_e \gets RJ$.get\_elastic\_jobs()
            \State $\mathcal{J}^r_e$.add$(\mathcal{J}^*.$get\_elastic\_jobs$())$
            \State $\mathcal{J}^*_e \gets$ AllocateElastic$(\mathcal{J}^r_e,
            {C}^*)$ \Comment{allocate for flexible demand}
            \State \textbf{return} $\mathcal{J}^*, \mathcal{J}^*_e$
        \EndIf
      \EndProcedure
  \end{algorithmic}
\end{algorithm}

\section{Jobs on Inference Servers}
\label{app:v100tot4}
In~\cref{sec:loaning}, we note that up to 21\% of jobs in our production trace
do not request specific GPUs. As long as the models and intermediate data fit into
the GPU memory, it can be trained on either training GPU or inference GPU.
Since inference GPUs have smaller memory of 16GB, some of these jobs need adjustment
of the batch size.
When \sys schedules these jobs on inference servers, it cut down the batch size
to half based on the theoretical GPU memory difference between V100 and T4
to ensure feasibility, if necessary.
\begin{table}[t]
  \centering
  \resizebox{\columnwidth}{!}{
  \begin{tabular}{@{}llllllllc@{}}
  \toprule
  \multirow{2}{*}{\#} & \multirow{2}{*}{Solution} & \multicolumn{3}{c}{Queuing Time} & \multicolumn{3}{c}{JCT}   & Preemption \\ \cmidrule(l){3-9}
                      &                           & Mean    & Median    & 95\%ile   & Mean  & Median & 95\%ile & Ratio      \\ \midrule
  1                   & Static                       & 3649    & 41        & 4993       & 18747 & 734    & 66524    & 36.75\%    \\
  2                   & SSF                       & 2354    & 26        & 3657       & 14953 & 688    & 62293    & 14.34\%    \\
  3                   & LSF                       & 2993    & 28        & 4774       & 12953 & 674    & 61005    & 28.58\%    \\
  4                   & \sys                      & 2204    & 23        & 3418       & 12414 & 655    & 57982    & 12.34\%    \\ \midrule
  % 4                   & \multirow{3}{*}{Basic}            & SSF                       & 2369    & 28        & 4602       & 11948 & 601    & 59832    & 27.49\%    \\
  % 5                   &                                   & LSF                       & 2231    & 25        & 3545       & 12556 & 641    & 60533    & 13.82\%    \\
  % 6                   &                                   & \sys                      & 2008    & 25        & 3356       & 11089 & 567    & 56477    & 10.20\%    \\ \bottomrule
  \end{tabular}
  }
  \vspace{-4mm}
  \caption{Simulation results using different reclaiming heuristics}
  \vspace{-2mm}
  \label{table:loaning_add_eval}
  \end{table}

\section{Existing Job Schedulers}
\label{app:scheduler_detail_comparison}
Apart from the discussion in \cref{sec:comparison}, we compare the \sys's
job scheduler with other existing schedulers in an algorithmic perspective. Table~\ref{table:elastic_comparison}
provides a summary.

\noindent\textbf{Dynamic capacity. }
\sys takes cluster capacity as a changing variable in both resource allocation
and job placement. During resource allocation, existing schedulers greedily allocate
resources job-by-job based on certain metrics regardless of the cluster capacity.
Specifically, Tiresias adopts the least attained
service to rank jobs; Optimus uses the largest marginal gains; AFS computes the
largest throughput improvement. In designing the scheduling policy,
these works ignore the outstanding question of whether there are remaining
resources to host them.
Though Pollux implicitly considers the cluster capacity during scheduling,
the random crossover in its genetic algorithm could easily violate the capacity
constraints, and repairing operation is randomly conducted.
\sys considers all possible allocations of each job and the cluster capacity
and groups the flexible workers during placement when the cluster
resource is dynamic. Though Gandiva supports flexible
system primitives to handle dynamic capacity, it adopts an opportunistic approach
to schedule jobs.

\noindent\textbf{Limited elasticity. }
\sys's job scheduler takes every elastic job demand into account during
allocation and effectively avoid excessive scaling.
Tiresias allocates resources based on the fixed job requirement.
Gandiva scales up the job when the cluster is underutilized and the
hosting servers have available resources.
Though it avoids excessive scaling, it neglects the scalability of the jobs.
AFS unlimitedly scales up jobs as long as they have a good throughput gain.
Pollux's scheduling policy states that a job can only be allocated with twice of
the maximum resources it has been allocated previously.
Eventually, it could still cause unlimited scaling. Optimus has no specific
boundaries for jobs but heavily relies on an accurate prediction of loss
convergence to adjust job resources.

\begin{table}[t]
  \centering
  \resizebox{\columnwidth}{!}{
\begin{tabular}{@{}lcccc@{}}
\toprule
                                & Dynamic capacity    & Limited elasticity &  Avoid starvation & Worker-unit scaling \\ \midrule
Tiresias\cite{gu2019tiresias}   &       \xmark       &     \xmark        &    \cmark           &       \xmark        \\
Optimus\cite{peng2018optimus}   &       \xmark       &     $\Delta$      &    \xmark           &       \cmark         \\
Gandiva\cite{xiao2018gandiva}   &       \xmark       &     $\Delta$      &    \xmark           &       \xmark         \\
AFS\cite{265013}                &       \xmark       &     \xmark        &    \cmark           &       \xmark         \\
Pollux\cite{qiao2020pollux}     &       $\Delta$     &     $\Delta$      &    \xmark           &       \xmark         \\
\sys                            &       \cmark       &     \cmark        &    \cmark           &       \cmark         \\ \bottomrule
\end{tabular}}
\vspace{-2mm}
\caption{Comparison with existing job schedulers. $\Delta$ indicates that it is handled implicitly.}
\vspace{-4mm}
\label{table:elastic_comparison}
\end{table}
\noindent\textbf{Avoid starvation. }
\sys allocates base demand of all queuing jobs in phase one to minimize
starvation, as well as for AFS.
Tiresias is a preemptive scheduler where starvation is handled properly with
priority promotion.
Optimus, Gandiva and Pollux do not launch as many jobs as possible initially,
incurring starvation for jobs at the end of the queue.

\noindent\textbf{Worker-unit scaling. }
\sys allocates resources in worker demand instead of GPUs.
It frees DL frameworks from adjusting the job's distributed architecture and
preserves the balance of pace each worker trains.
Only Optimus considers the parameter server architecture of distributed jobs
and adjust the workers of each job.
AFS naively allocates 1 GPU at a time; Pollux only cares about how many GPUs
should be allocated to each job. They are built on an assumption
that a running job could either switch between different training modes
or balance the training pace with negligible effort. Both are still
operations requiring delicate adjustments.

\begin{comment}
\noindent\textbf{Modeling of training performance. }
Pollux and Optimus need online fitting to profile job performance and make
scheduling decisions. The impracticality lies in the requirement of accurate
profiling and the heavy cost of collecting necessary data to profile.
For instance, Pollux needs to learn seven parameters to compute the goodput
of a job, let alone its assumption on the ideal scenario. Moreover, one has to collect
lots of data beforehand to obtain accurate results. In a heavy-loaded cluster
where low-overhead scheduling is required, the efficiency cannot be guaranteed.
\end{comment}